\documentclass[journal]{IEEEtran}
\usepackage{mathtools,amssymb,bm}
\usepackage{amsmath}
\usepackage{cite}
\usepackage{graphicx}
\usepackage{epstopdf}
\usepackage{cite}
\usepackage{color}
\usepackage{caption}
\usepackage[footnotesize]{subfigure}
\usepackage{hyperref}
\newcommand{\tens}[1]{%
	\mathbin{\mathop{\otimes}\limits_{#1}}%
}
\newtheorem{thm}{Theorem}[section]
\newtheorem{lem}[thm]{Lemma}
\newtheorem{prop}[thm]{Proposition}

\newtheorem{defn}{Definition}[section]

\begin{document}
	\title{Two-Dimensional Super-Resolution via Convex Relaxation}
	\author{Iman Valiulahi, Sajad Daei, Farzan Haddadi and Farzad Parvaresh
}
	\maketitle
	\begin{abstract}
		In this paper, we address the problem of recovering point sources from two dimensional low-pass measurements, which is known as super-resolution problem. This is the fundamental concern of many applications such as electronic imaging, optics, microscopy, and line spectral estimation. We assume that the point sources are located in the square $[0,1]^2$ with unknown locations and complex amplitudes. The only available information is low-pass Fourier measurements band-limited to integer square $[-f_c,f_c]^2$.  The signal is estimated by minimizing Total Variation $(\mathrm{TV})$ norm, which leads to a convex optimization problem. It is shown that if the sources are separated by at least $1.68/f_c$, there exist a dual certificate that is sufficient for exact recovery.
		\end{abstract}
\section{Introduction}
Two dimensional (2-D) super resolution refers to recovering 2-D point sources from their low-resolution measurements. One may think of this problem as recovering a high resolution image of stars in a photo captured by a low-resolution telescope. Various other fields are also involved in this problem. In the direction of arrival (DOA) estimation, far field point sources are to be located in terms of their 2-D directions using measurements by a two-dimensionally dispersed array of sensors. Applications are vast from Radar, sonar to cellular communication systems. The main performance measure for any DOA estimation algorithm is its ability to resolve two closely-spaced sources which leads to the term super-resolution methods \cite{krim1996two}. High-dimensional super-resolution has also important applications in off-the-grid Multiple-input multiple-output (MIMO) radar where the aim is to estimate the angle-delay-Doppler continuous triplets from the reflections recorded at receiver antennas\cite{hecket2016generalized}.
Another example is high dimensional medical imaging, a diagnosis method to determine the presence of some certain diseases\cite{greenspan2008super}. \par Consider $r$ high-frequency 2-D signals in the form of Dirac delta functions. The signal is observed after convolution with a low-pass kernel. In some applications, there is no exact information about the kernel. In this case, joint estimation of the signal and the kernel is required which is blind super-resolution \cite{campisi2016blind}, \cite{chi2016guaranteed}. In many scenarios, the low-pass kernel is known before-hand. In this paper, we assume that the signal is observed through convolution with a 2-D sinc kernel band-limited to the integer square $[-f_c,f_c]^2$. With this assumption, the measurements are in the form of superposition of $r$ sinusoids with arbitrary complex amplitudes. This model has a closed relation with 2-D line spectral estimation.\par
Conventional parametric approaches to super-resolve sparse 2-D point sources are based on decomposition of measurement space into orthogonal signal and noise subspaces such as 2-D MUSIC \cite{hua1993pencil}, 2-D unitary ESPRIT \cite{haardt19952d} and Matrix Enhancement Matrix Pencil (MEMP) method\cite{hua1992estimating}. However, these techniques are sensitive to noise and outliers. They are also dependent on model order. Discrete 2-D super-resolution suggests that one can recover the sparse signal by solving an $\ell_1$ minimization problem \cite{santosa1986linear,levy1981reconstruction,baraniuk2007compressive,candes2006robust}. This method assumes all the point sources to lie on the grid. However, this assumption is not realistic in practice. When the true point sources do not lie on the grid, basis mismatch occurs which leads to reduced performance. One is able to achieve better reconstruction using finer grids, but this imposes higher computational complexity \cite{duarte2013spectral,chi2011sensitivity}. To overcome grid mismatch, \cite{candes2014towards} presented a new method based on convex optimization that recovers the infinite-dimensional signal from low-resolution measurements by minimizing a continuous version of $\ell_1$ norm known as Total Variation ($\mathrm{TV}$) norm. Similar to compressed sensing, a sufficient condition for exact recovery is the existence of a dual certificate orthogonal to the null-space of the measurements with sign pattern of the signal in the support and magnitude less than one in off-support locations. \cite{candes2014towards} constructed this dual certificate as a linear combination of $r$ shift copies of forth power of Dirichlet kernel (and its derivatives). They prove that existence of such a linear combination imposes $2/f_c$ minimum separation between the point sources for 1-D situation where $f_c$ is the cut-off frequency. In 2-D case, the sources must be separated at least by $2.38/f_c$ to construct dual polynomial band-limited to integer square $[-f_c,f_c]^2$.\par 
 Implementation of $\mathrm{TV}$ norm minimization problem may seem tough because of infinite dimensionality of the primal variable. To handle this situation, one can convert the dual problem to a tractable semidefinite program (SDP) using Positive Trigonometric Polynomial (PTP) theory. In fact, PTP theory provides conditions to control the magnitude of trigonometric polynomials in signal domain by some linear matrix inequalities (LMI) \cite{dumitrescu2007positive,xu2014precise,chi2015compressive}. Moreover, it is possible to control the magnitude of trigonometric polynomial in any partition of signal domain which can be translated to prior information \cite{mishra2015spectral,valiulahi2017off}.\par 
The approach of \cite{candes2014towards} was extended to off-the-gird spectral estimation in compressed sensing (CS) regime\cite{tang2013compressed}. It shows that atomic norm minimization can recover a 1-D continuous spectrally sparse signal from partial time domain samples as long as the frequency sources are separated by $4/n$ where $n$ is number of Nyquist samples. Proof is based on constructing a random dual certificate that guarantee exact recovery with high probability. Similar to this work, \cite{chi2015compressive} presented 2-D random dual polynomial time-limited to integer square $[-2M,2M]^2$ for estimating the true off-the-grid 2-D frequencies under the condition that they satisfy a minimum separation $1.19/M$. \cite{chen2013spectral} investigated multi-dimension frequency reconstruction by minimizing nuclear norm of a Hankel matrix subject to some time domain constraints known as Enhanced Matrix Completion (EMaC).
Recently, Fernandez in \cite{fernandez2016super}, has shown that the guaranty for exact recovery of 1-D point sources using $\mathrm{TV}$ norm minimization can be improved up until the minimum separation $1.26/f_c$ by constructing a dual certificate that interpolates the sign pattern of closer point sources. The main idea of his work is to use the product of $p$ Dirichlet kernels with different bandwidths instead of its forth power that was used in \cite{candes2014towards} for 1-D case. This leads to a better trade-off between the spikiness of the dual polynomial in the support locations and decay of its tail.\par
The main result of this paper is to guarantee that $\mathrm{TV}$ norm minimization achieves exact solution as long as the 2-D sources are separated by at least $1.68/f_c$. Specifically, we extend the approach of Fernandez to the recovery of 2-D point sources which lie in $[0,1]^2$ with arbitrary locations and complex amplitudes. For this purpose, we first propose a 2-D low-pass kernel caped by the integer square $[-f_c,f_c]^2$ which is obtained by tensorizing the 1-D kernel used by Fernandez \cite{fernandez2016super}. In comparison with the 2-D kernel used by \cite{candes2014towards}, our kernel better balances between spikiness at the origin and decay of its tail. Then we construct 2-D low-pass dual polynomial by linearly combining $r$ shifted copies of the kernel and its partial derivatives. Our theoretical guaranty requires bounds on the 1-D and 2-D kernels which are verified by numerical simulations given in Section \ref{sec.exprement}. \\The rest of the paper is organized as follows. The problem is formulated in Section \ref{sec.formulation}. Section \ref{sec.TV} presents $\mathrm{TV}$ norm minimization and the proposed uniqueness guaranty. Implementation of the dual problem is given in Section \ref{sec.imple}. In Section \ref{sec.exprement} results are validated via numerical experiments. Finally, we conclude the paper and introduce future directions in Section \ref{conclude}.\par
\textbf{Notation}.
 Throughout the paper, scalars are denoted by lowercase letters, vectors by lowercase boldface letters, and matrices by uppercase boldface letters. The $i$th element of the vector $\bm{x}$ and the $\bm{k}=(k1,k2)$ element of the matrix $\bm{X}$ are given by $x_i$ and $x_{\bm{k}}$, respectively. $|\cdot|$ denotes cardinality for sets and absolute value for scalars. For a function $f$ and a matrix $\bm{A}$, $\|f\|_{\infty}$ and $\|\bm{A}\|_{\infty}$ are defined as $\|f\|_{\infty}=\underset{t}{\sup}|f(t)|$ and $\|\bm{A}\|_{\infty}=\underset{\|\bm{x}\|_{\infty}\le1}{\sup}\|\bm{Ax}\|_{\infty}=\max_i\sum_j|A_{i,j}|$, respectively. Null space of linear operators are denoted by $\mathrm{null}(\cdot)$. $\mathrm{relint}(C)$ denotes relative interior of a set $C$. $f^{(i}(t)$ and $f^{i_1,i_2}(\bm{t})$ denote $i$th derivate and $i_1,i_2$ partial derivatives of 1-D function $f(t)$ and 2-D function $f(\bm{t}:=(t_1,t_2))$, respectively. $(\cdot)^T$ and $(\cdot)^{H}$ shows transpose and  hermitian of  a vector, respectively. $\text{sgn}(\bm{x})$ denotes the element-wise sign of the vector $\bm{x}$. Also, $\mathrm{vec}(\bm{X})$ denotes the columns of $\bm{X}$ being stacked on top of 
 each other. The inner product between two functions $f$ and $g$ is defined as $\langle f,g \rangle:\int f(t) g(t) dt$. $\tens{C}$ and $\tens{D}$ denotes tensor and Kronecker product, respectively. The adjoint of a linear operator $\mathcal{F}$ is denoted by $\mathcal{F}^*$.   
\section{Problem Formulation}\label{sec.formulation}
We consider a mixture of $r$ 2-D Dirac function on a continuous support $T$:
\begin{align}
\label{eq1}
\bm{x}_{2D}(\bm{t})=\sum_{i=1}^{r} d_i \delta(\bm{t}-{\bm{t}_i}),
\end{align}
where $d_i=|d_i|e^{j\phi_i}$ is an arbitrary complex amplitude of each point source with $\phi_i \in [0,2\pi)$, $\bm{t}_i:=[t_{1i},t_{2i}]^T$ in the continuous square $[0,1]^{2}$, and $\delta(\cdot)$ denotes Dirac function. Assume that the only available information about $\bm{x}_{2D}$ is its 2-D Fourier transform band-limited to integer square $[-f_c,f_c]^2$ as:
\begin{align}
\label{eq2}
y_{\bm{k}}=\int_{[0,1]^{2}} e^{-j2\pi\langle\bm{k},\bm{t}\rangle}\bm{x}_{2D}(\bm{t})(d\bm{t})=\sum_{i=1}^{r} d_i~ e^{-j2\pi\langle\bm{k},\bm{t}_j\rangle},
\end{align}
where $\bm{k}=(k_1,k_2)\in J$,  $J=\{-f_c, . . . , f_c\}\times\{-f_c, . . . ,f_c\}$ denotes all of indices of the signal ($f_c$ is an integer). It is beneficial to consider each observation an element of a matrix $\bm{Y} \in\mathbb{C}^{n\times n}$ as below:
\begin{align}
\label{eq3}
\bm{Y}:=\mathcal{F}_{2D}\bm{x}_{2D},
\end{align}
which $n:=2f_c+1$ and $\mathcal{F}_{2D}$ is the 2-D linear operator that maps a continuous function to its lowest 2-D Fourier coefficients up until the integer square $[-f_c,f_c]^2$. The problem is then to estimate $\bm{x}_{2D}$ from the observation matrix $\bm{Y}$. 
\section{Total Variation Minimization for 2-D Sources}\label{sec.TV}
To super-resolve the point sources from Fourier measurements, one can use the following optimization problem: 
\begin{align}
\label{eq4}
\mathrm{P}_{\mathrm{TV}}:~~~\min_{\bm{z}_{2D}}~\|\bm{z}_{2D} \|_{\mathrm{TV}}~~\mathrm{subject\,\, to}~~ \bm{Y}=\mathcal{F}_{2D}\bm{z}_{2D},
\end{align}
where $\mathrm{TV}$ norm promotes sparse atomic measures which define as:
\begin{align}\label{eq.TVnormdef}
\|\bm{z}_{2D}\|_{\mathrm{TV}}:=\sup_{\rho}\sum_{E\in\rho}|\bm{z}_{2D}(E)|,
\end{align}
in which $\rho$ is any partition of $[0,1]^2$ into finite number of disjoint measurable 2-D subsets and $|\bm{z}_{2D}(E)|$ is a positive measure on $E$. In particular, $\|\bm{x}_{2D}\|_{\mathrm{TV}}=\sum_{i=1}^r|d_i|$.\\The main goal of this paper is to show that, $\mathrm{P}_{\mathrm{TV}}$ exactly recovers $\bm{x}_{2D}$ if the sources satisfy some mild separation. In the following, we define the minimum distance of a point from a 2-D set. 
\begin{defn} Let $\mathbb{T}^{2}$ be the product space of two circles obtained by identifying the endpoints on $[0,1]^2$. For each set of points $T\subset\mathbb{T}^2$, the minimum separation is defined as:
	\begin{align}
	\label{eq6}
	&\Delta(T):=\underset{\bm{t}_i,\bm{t}_j\in T,~ \bm{t}_i\neq \bm{t}_j}\inf~\|\bm{t}_i-\bm{t}_j \|_\infty\nonumber\\
	&=\underset{\bm{t}_i,\bm{t}_j\in T,~ \bm{t}_i\neq \bm{t}_j} \inf\max \{|t_{1i}-t_{1j}|,|t_{2i}-t_{2j}|\},
	\end{align}
	where $|t_{1i}-t_{1j}|$ and $|t_{2i}-t_{2j}|$ denote warp-around distances on the unit circle.
\end{defn}
It has been shown that $\mathrm{P}_{\mathrm{TV}}$ can achieve exact recovery for $f_c\ge 512$ as long as the components of the support are separated by at least $2.38/f_c$ \cite{candes2014towards}. The following theorem which is the main result of this paper states that for $f_c\ge2\times10^3$ under some milder separation condition on the support, $\mathrm{P}_{\mathrm{TV}}$ can reach the exact solution.
\begin{thm}\label{thm.main}
Let $T=\{\bm{t}_j\}_{j=1}^r$ be the support of $\bm{x}_{2D}$. If $f_c \geq 2\times 10^{3}$ and the minimum septation obeys 
\begin{align}
\label{eq7}
\Delta(T) \geq 1.68\lambda_c,
\end{align}
 where $\lambda_c:=1/f_c$, then the solution of $\mathrm{P}_{\mathrm{TV}}$ is unique.  
\end{thm}
\subsection{Uniqueness Guaranty}
To prove uniqueness of $\mathrm{P}_{\mathrm{TV}}$, it is sufficient to find a function $Q(\bm{t})$ known as dual certificate which is orthogonal to the null space of $\mathcal{F}_{2D}$ and belongs to relative interior of sub-differential of $\mathrm{TV}$ norm at the original point $\bm{x}_{2D}$\cite{candes2014towards}.
\begin{prop}\label{prop.main}
If the conditions of Theorem \ref{thm.main} hold, then for any sign pattern $\bm{v}\in \mathbb{C}^{|T|}$ with $|v_j|=1~\forall j$ there exist a low-pass function
\begin{align}
\label{eq8}
Q(\bm{t})=\,\sum_{\bm{k}\in J}q_{\bm{k}} e^{j2\pi \langle\bm{t},\bm{k}\rangle},
\end{align}
such that 
\begin{align}
&Q(\bm{t}_i)=v_i, && \bm{t}_i \in T,\label{cond1}\\
&|Q(\bm{t})| <1, &&\bm{t} \notin T\label{cond2}.
\end{align}
\end{prop}
 The conditions (\ref{eq8}),  (\ref{cond1}) and (\ref{cond2}) refer to  the fact that $Q(\bm{t})\in \mathrm{null}^{\perp}(\mathcal{F}_{2D}) ~\cap~ \mathrm{relint}(\partial\|\cdot\|_{\mathrm{TV}}(\bm{x}_{2D}))$. The proof of Proposition is given in Appendix \ref{proofprop}.
\\It is beneficial to emphasize that if the sign of closely-spaced sources differ from each others, then it is ill-posed to interpolate sign pattern of ${\bm{x}_{2D}}$ a the low-pass trigonometric polynomial. That's why the minimum separation condition is required. This can be rebated if the sign of sources be the same as in \cite{morgenshtern2014stable,schiebinger2015superresolution}.
\subsection{Construction of the dual certificate}
To construct the dual certificate $Q(\bm{t})$ in Proposition \ref{prop.main} under the conditions of Theorem \ref{thm.main}, we first propose the following 2-D low-pass kernel:
\begin{align}
\label{eq12}
&K_{2D}=K_{\bm{\gamma}}\tens{C}K_{\bm{\gamma}},
\end{align}
where $K_{\bm{\gamma}}(t),\quad\forall t \in [0,1]$ is multiplication of three Dirichlet kernel $K(f,t)$ with different cut-off frequencies defined as below:
\begin{align}
\label{eq10}
K_{\bm{\gamma}}(t)=\prod_{i=1}^{3}K(\gamma_if_c,t)=\sum_{k=-f_c}^{f_c}c_ke^{j2\pi kt},
\end{align}
where
\begin{align}\label{Dirch}
K(f,t)=\frac{1}{2f+1}\sum_{k=-f}^{f}e^{j2\pi kt},
\end{align}
in which $f$ is the cut-off frequency, $\gamma_1=0.247$, $\gamma_2=0.339$, $\gamma_3=0.414$, and $\bm{c} \in \mathbb{C}^n$ is the convolution of the Fourier coefficient of $K(\gamma_1f_c,t)$, $K(\gamma_2f_c,t)$, and $K(\gamma_3f_c,t)$. Consequently,
\begin{align}\label{twod}
&K_{2D}(\bm{t})=\sum_{\bm{k}\in J}c_{k_1}c_{k_2}e^{j2\pi \langle\bm{t},\bm{k}\rangle}.
\end{align}
 Fernandez in \cite{fernandez2016super} proposed $K_{\bm{\gamma}}(t)$ for 1-D situation instead of forth power of Dirichlet kernel that was previously used in \cite{candes2014towards}. This kernel provides a better trade-off between spikiness in the origin and the order of tail decay. This motivates proposition of 2-D kernel in the form of (\ref{eq12}).\par
If $Q(\bm{t})$ was constructed such that only (\ref{cond1}) is satisfied, the magnitude of the resulting polynomial may exceed one near the elements of the support $T$. To handle this situation, we force the derivative of the polynomial to be zero at the support of $\bm{x}_{2D}$. We construct $Q(\bm{t})$ as
\begin{align}
\label{eq13}
Q(\bm{t})=\sum_{\bm{t}_i \in T}\alpha_iK_{2D}(\bm{t}-\bm{t}_i)+\nonumber\beta_{1i}K_{2D}^{10}(\bm{t}-\bm{t}_i)\\
+\beta_{2i}K_{2D}^{01}(\bm{t}-\bm{t}_i),
\end{align}
to better control $K_{2D}(\bm{t})$ and its derivatives. $K_{2D}^{10}(\bm{t})$ and $K_{2D}^{01}(\bm{t})$ denote the partial derivatives of $K_{2D}(\bm{t})$ with respect to $t_1$, $t_2$, respectively. Therefore, instead of (\ref{cond1}) and (\ref{cond2}), the following conditions are considered.
\begin{align}
&Q(\bm{t}_i)=v_i,&&\bm{t}_i\in T,\label{signcondition}\\
&\nabla Q(\bm{t}_i)=0,&&\bm{t}_i\in T\label{diffcondition}.
\end{align}
In Appendix \ref{proofprop}, it is shown that one can always find interpolation coefficients $\bm{\alpha},\bm{\beta}_1,\bm{\beta}_2\in\mathbb{C}^{|T|}$ under the conditions of Theorem \ref{thm.main}.
  \begin{figure*}[t]
	\centering
	\subfigure[]{\includegraphics[width=3.5in]{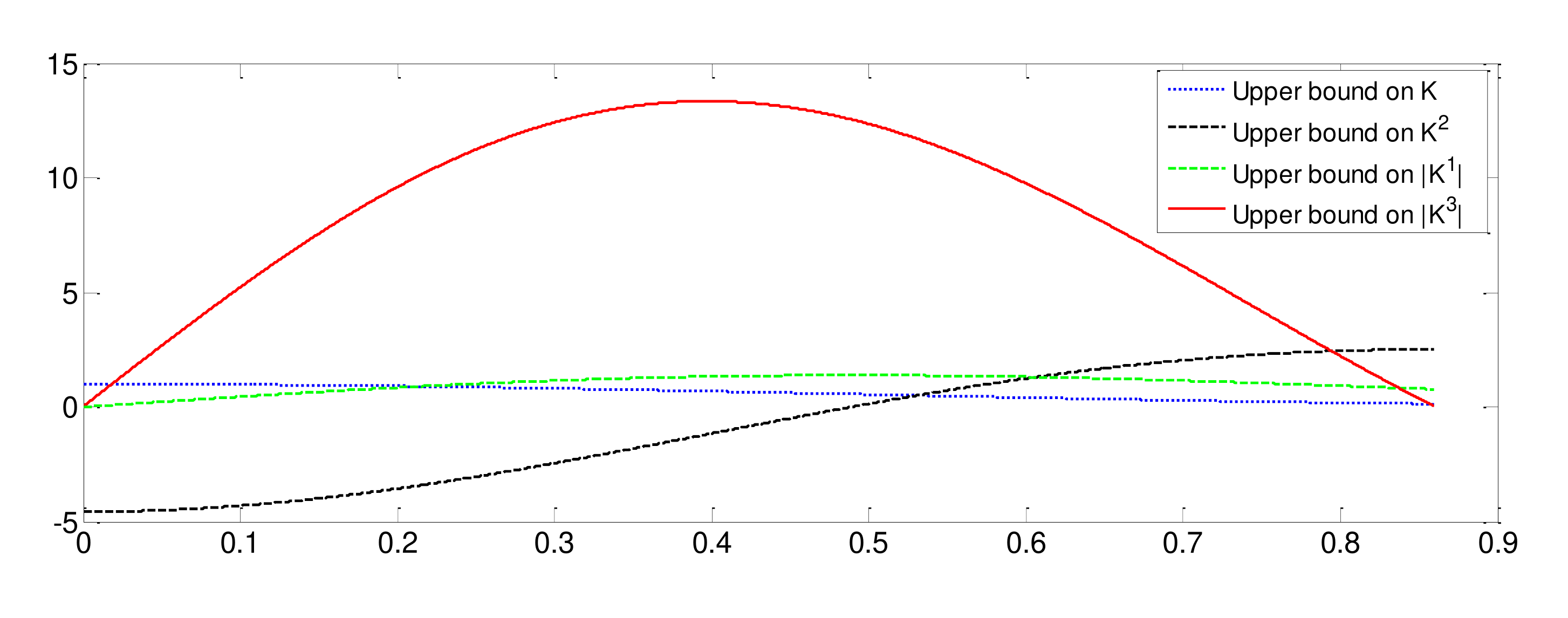}\label{fig.bound1}}
	
	\mbox{
		\subfigure[]{\includegraphics[width=3.5in]{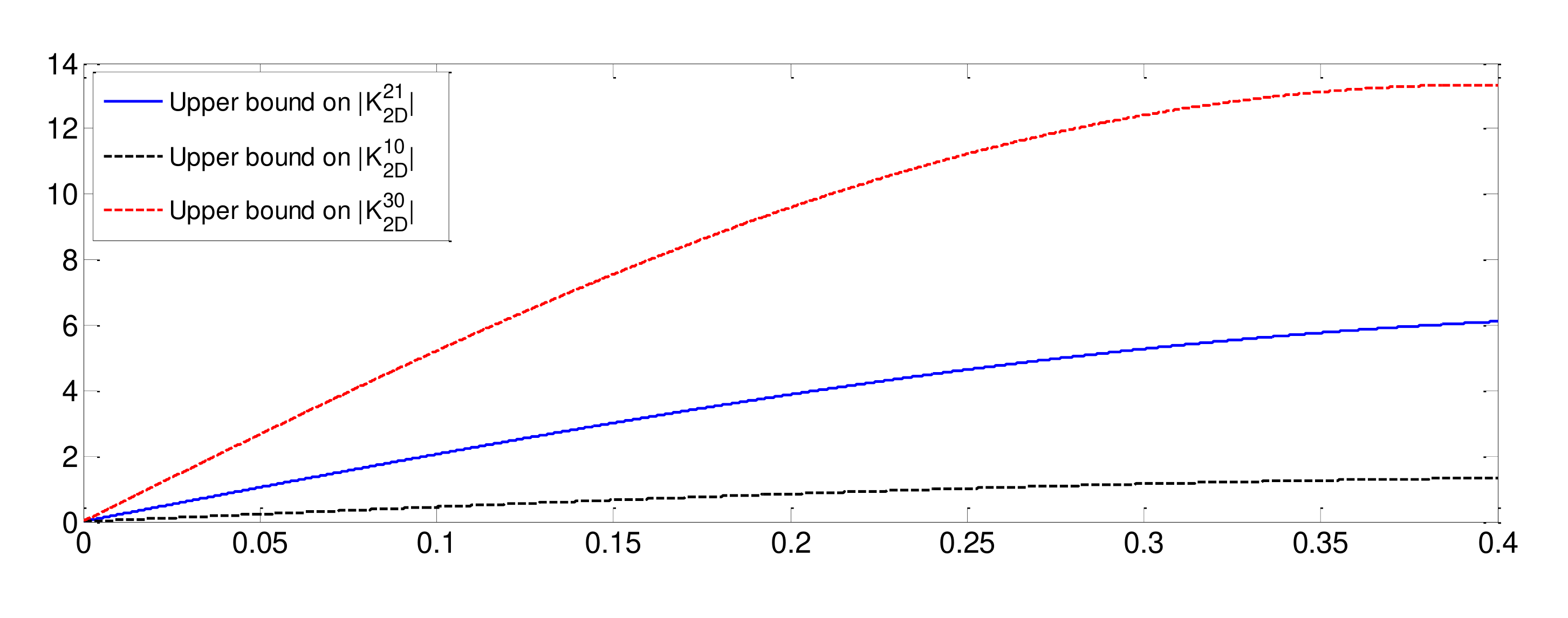}\label{fig.bound4}}\quad
		\subfigure[]{\includegraphics[width=3.5in]{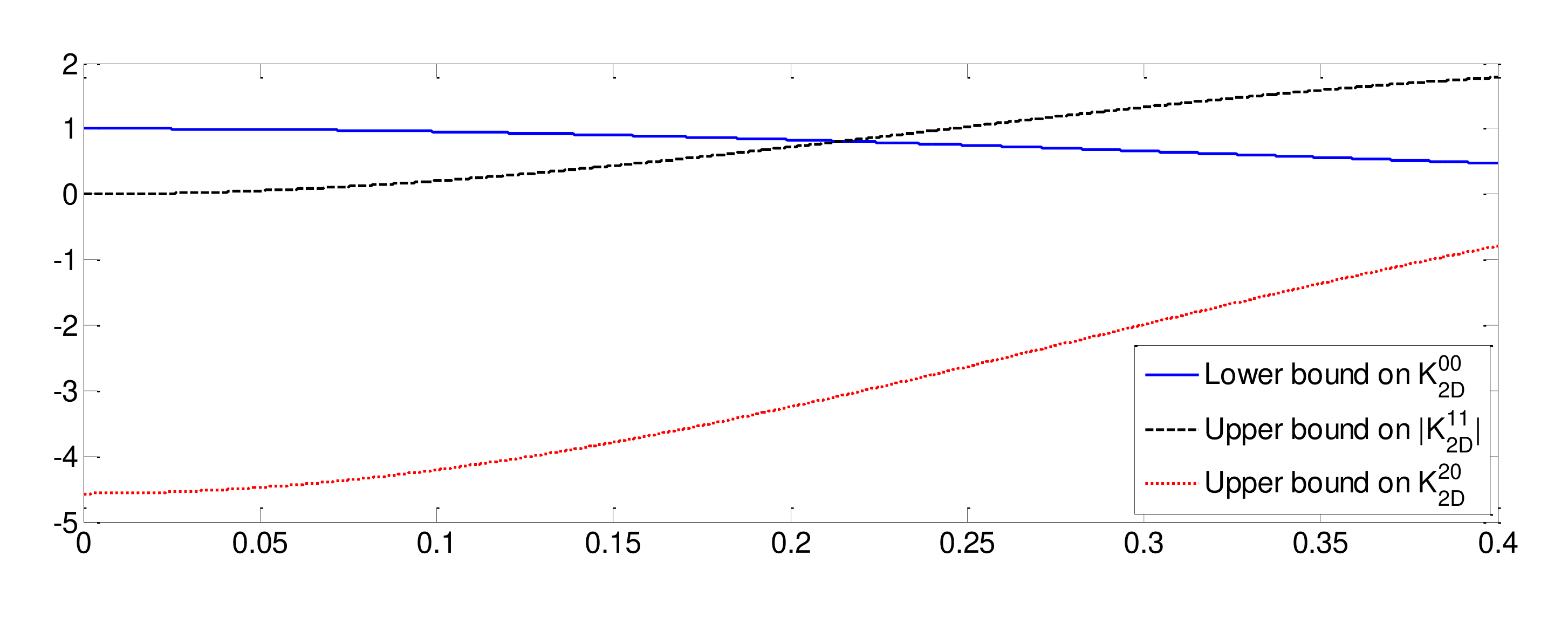}\label{fig.bound5}}}
	\caption{\subref{fig.bound1}Bounds on 1-D kernel $K_{\bm{\gamma}}^\ell(t)/f_c^\ell$. The bounds are calculated on the grid size $\epsilon=10^{-6}$ covering the interval $0\le t/\lambda_c\le1.68/2$. All bounds are monotone in $[0,0.212568]$. \protect \subref{fig.bound4} and \subref{fig.bound5} Bounds on $K_{2D}^{i_1 i_2}(\bm{t})/f_c^{i_1+i_2}$ and its partial derivatives. The bounds are calculated on the grid size $\epsilon=10^{-6}$ covering the interval $0\le t_1/\lambda_c\le0.4$. }\label{fig.boundk1k2}
\end{figure*}
\begin{figure*}[t]
	\centering
	\mbox{\subfigure[]{\includegraphics[width=3.5in]{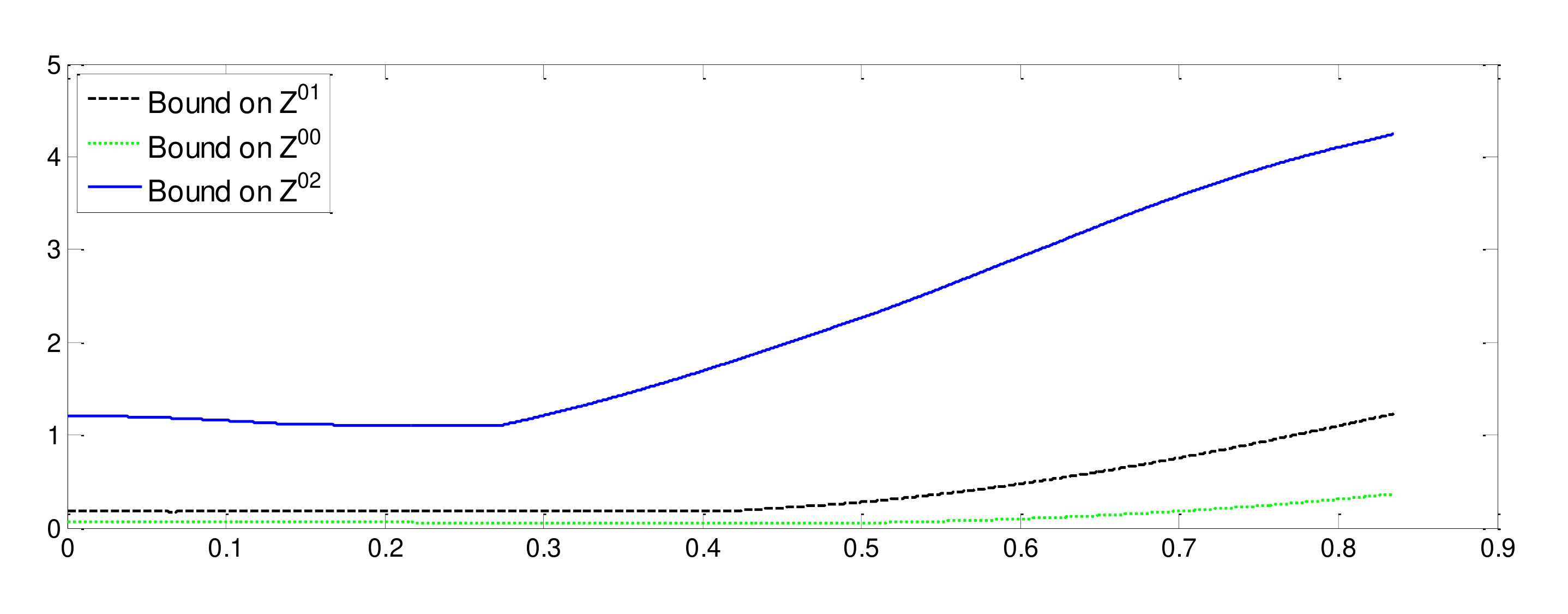}\label{fig.bound2}}\quad
		\subfigure[]{\includegraphics[width=3.5in]{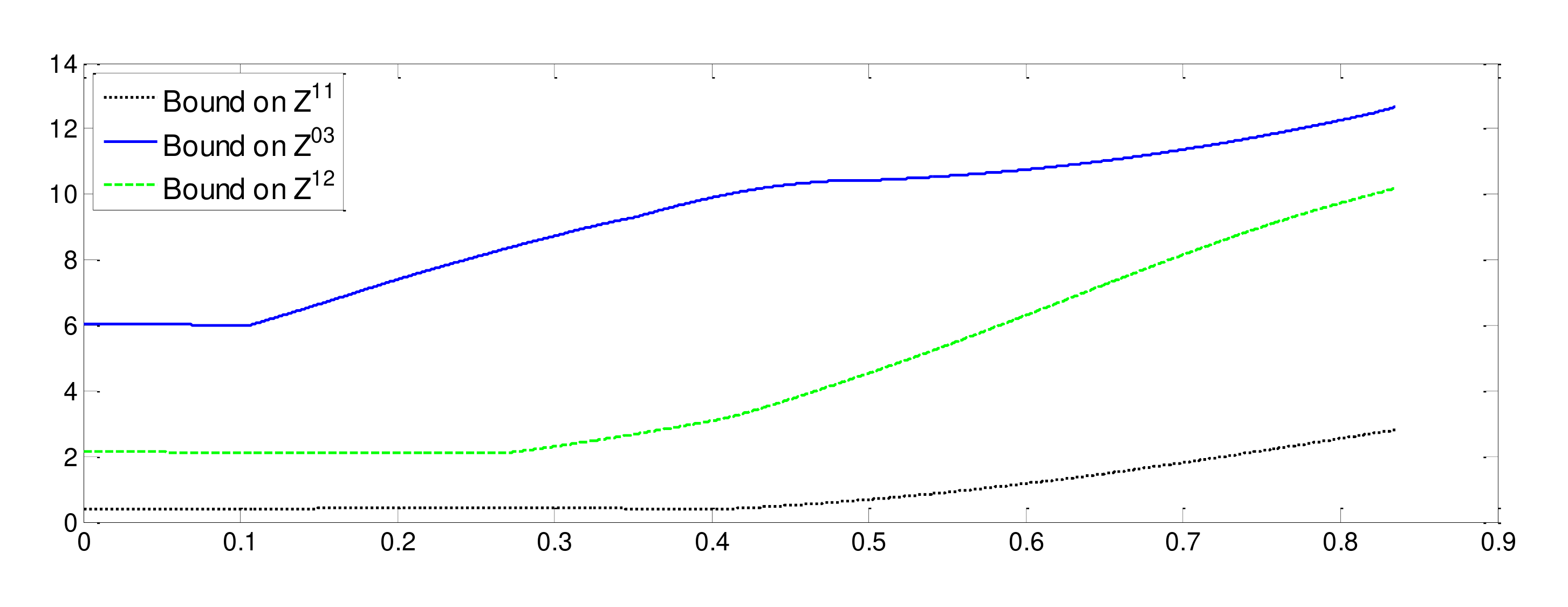}\label{fig.bound3}}}
	\caption{Bounds on $\bm{Z}^{i_1i_2}(\|\bm{t}\|_2)/f_c^{i_1+i_2}$ in (\ref{eq47}). The bounds are calculated on the grid size $\epsilon=10^{-6}$ covering the interval $0\le\|\bm{t}\|_2/\lambda_c\le 1.68/2$.}\label{fig.boundZ}
\end{figure*}
\begin{figure*}[!t]
	\centering
	\includegraphics[scale=.5]{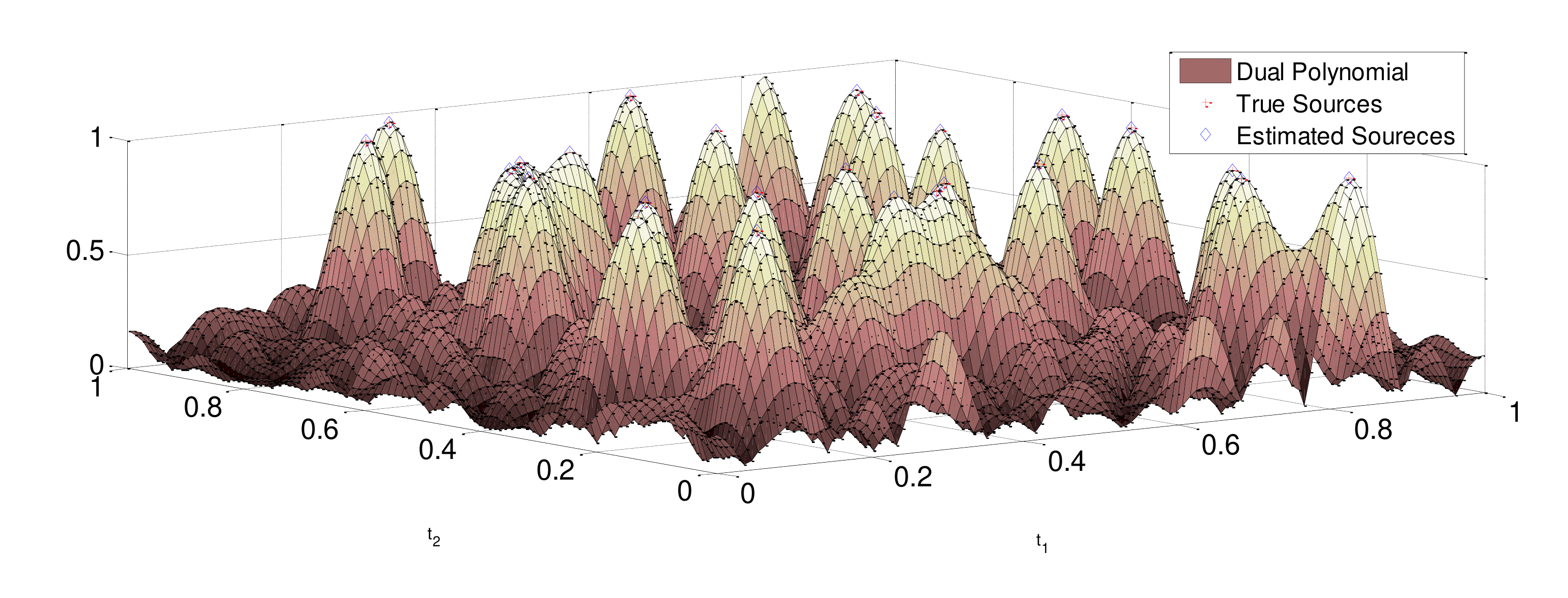}
	\caption{2-D Point sources recovery using the dual polynomial (\ref{eq.dualpolynomial}). The true sources are estimated by localizing the modulus of the dual polynomial is one. }\label{fig.dualpoly}
\end{figure*}
 \begin{figure*}[t!]
	\centering
	\mbox{\subfigure[]{\includegraphics[width=3.7in]{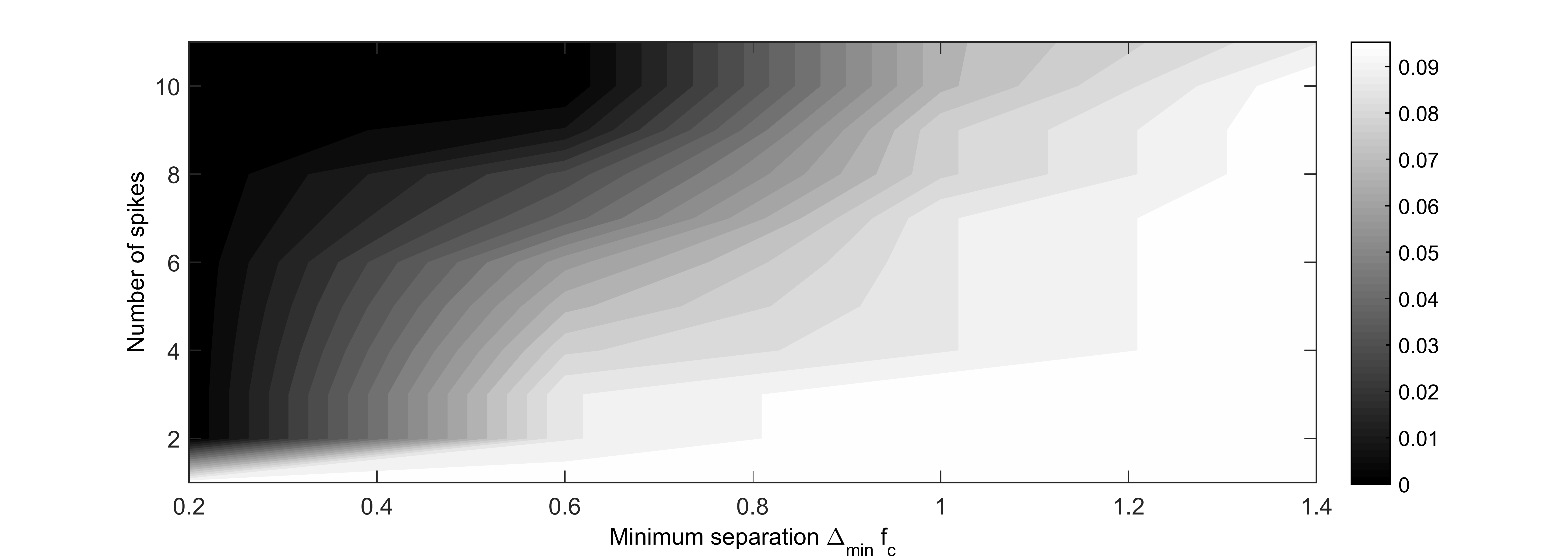}\label{fig.fc8}}\quad
		\subfigure[]{\includegraphics[width=3.7in]{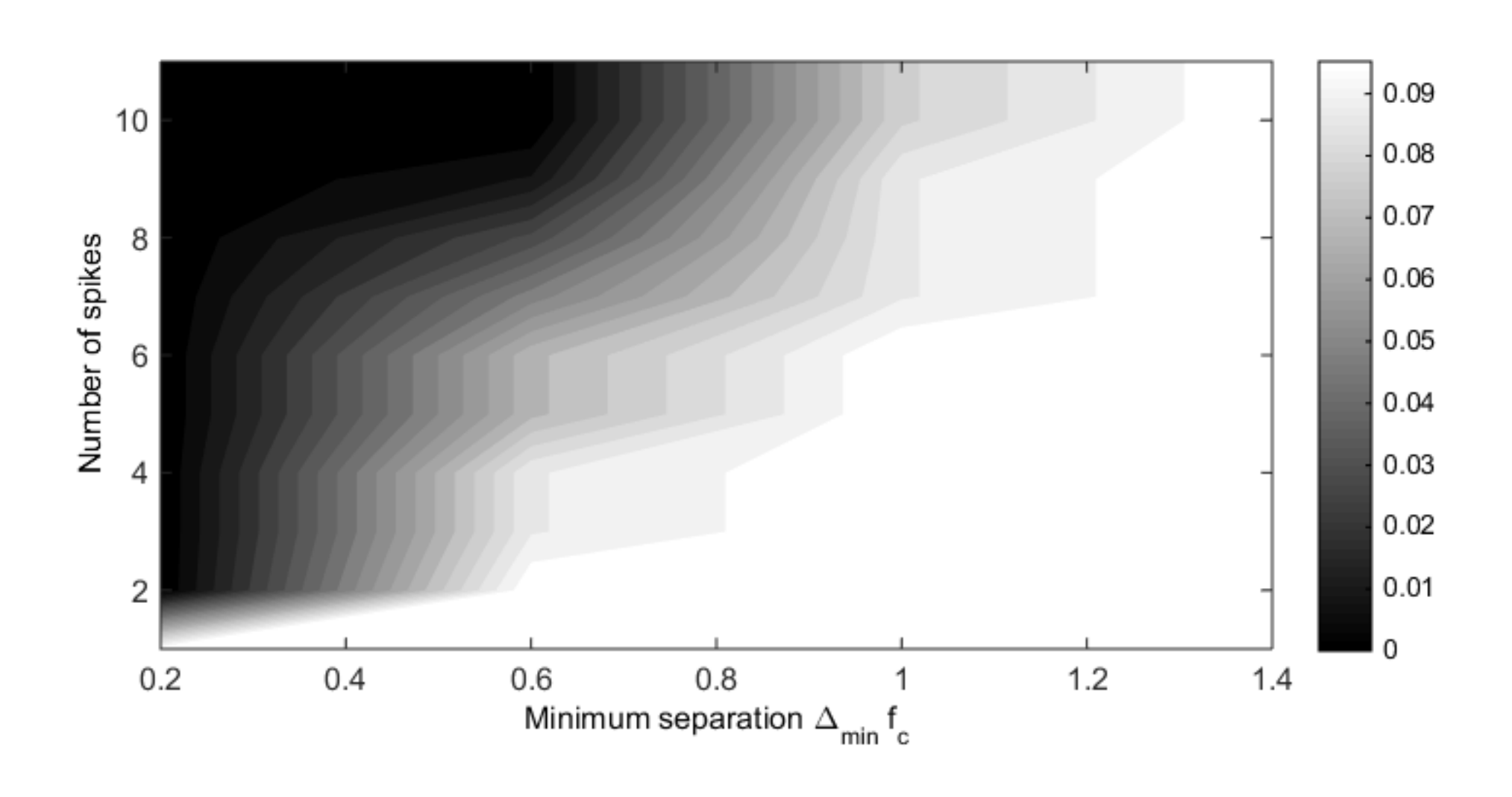}\label{fig.fc10}}}
	\caption{Graphs showing phase transition of successful recovery of $\mathrm{P}_{\mathrm{TV}}$ over 10 trials. In \subref{fig.fc8} and \protect \subref{fig.fc10}, the number of samples is $64$ and $100$, respectively.}\label{fig.phase}
\end{figure*}
\section{Implementation}\label{sec.imple}
It may seem challenging to find an exact solution of $\mathrm{P}_{\mathrm{TV}}$ since the variable lies in a continuous domain. Due to the establishment of Slater's condition and convexity of $\mathrm{P}_{\mathrm{TV}}$ strong duality holds. Therefore, one can consider the following dual problem
 \begin{align}
 \label{eq17}
 \max_{\bm{C}\in\mathbb{C}^{n\times n}}~\mathrm{Re}{\langle\bm{C},\bm{Y}\rangle_F}~~\mathrm{subject\,\, to}~~ \|\mathcal{F}_{2D}^{*}\bm{C}\|_{\infty} \le 1,
 \end{align}
 where $\mathrm{Re}{\langle\cdot\,,\cdot\rangle_F}$ denotes the real part of Frobenius inner product, $\bm{C}$ is the dual variable, and the inequality constraint implies that the modulus of the following trigonometric polynomial is uniformly bounded by $1$:
  \begin{align}
  (\mathcal{F}_{2D}^{*}\bm{C})(\bm{t}):=\sum_{\bm{k}\in J}^{}c_{\bm{k}} e^{j2\pi \langle\bm{t},\bm{k}\rangle}.
  \end{align}
  Also this inequality can be converted to some linear matrix inequalities using PTP theory. Then the problem can be considered as a SDP which can be solved in polynomial time \cite{xu2014precise}. Hence, (\ref{eq17}) is equivalent to
  \begin{align}\label{sdp}
  \begin{split}
  &\underset{\bm{C},\bm{Q}_0}{\max} ~~\mathrm{Re}{\langle \bm{Y},\bm{C}\rangle}_F\\
  &\mathrm {subject\,\,to }~ \delta_{\bm{k}}=\mathrm{tr}[\bm{\Theta}_{\bm{k}}\bm{Q}_0],\quad \bm{k}\,\in\ J,
  \end{split}
  \end{align}
  \[
  ~~\begin{bmatrix}
  \bm{Q}_0 & {\mathrm{vec}(\bm{C})} \\
  \\({\mathrm{vec}(\bm{C})})^H & \bm{1}
  \end{bmatrix}
  \succeq \bm{0},\
  \]
  where $\bm{Q}_0\in \mathbb{C}^{n^2\times n^2}$ is a positive semidefinite Hermitian matrix, $\bm{\Theta}_{\bm{k} }=\bm{\Theta}_{k_2 }\tens{D}\bm{\Theta}_{k_1}$, $\bm{\Theta}_k\in\mathbb{C}^{n\times n} $\,is an elementary Toeplitz matrix with ones on it's k-th diagonal and zeros else where. $\delta_{\bm{k}}=1$ if $\bm{k}=(0,0)$, and zero otherwise.\par
  By strong duality, for any solution $\hat{\bm{x}}_{2D}$ and $\hat{\bm{C}}$ of $\mathrm{P}_{\mathrm{TV}}$ and (\ref{sdp}), respectively, we have:
  \begin{align}\label{eq.howtofindsupport}
  &\langle \mathcal{F}_{2D}\bm{x}_{2D},\hat{\bm{C}}\rangle_{F}=\langle \bm{x}_{2D},\mathcal{F}_{2D}^*\hat{\bm{C}}\rangle=\|\hat{\bm{x}}_{2D}\|_{\mathrm{TV}}=\nonumber\\
  &\langle\hat{\bm{x}}_{2D},\mathrm{sgn}(\hat{\bm{x}}_{2D})\rangle.
  \end{align}
  Therefore, $(\mathcal{F}_{2D}^*\hat{\bm{C}})(\bm{t})=\mathrm{\mathrm{sgn}}(\hat{\bm{x}}_{2D}(\bm{t}))~\forall \bm{t}\in T$. This suggests that one can find the support by looking for $\bm{t}\in [0,1]^2$ such that  $|(\mathcal{F}_{2D}^*\hat{\bm{C}})(\bm{t})|=1$ (see Fig. \ref{fig.dualpoly}). 
\section{Experiment}\label{sec.exprement}
In this section, we numerically provide some bounds on $K_{\bm{\gamma}}(t)$ and its derivatives to show their monotonicity in the interval $|t|\le 0.212568\lambda_c$ which is necessary in Lemma \ref{lemasli2}. These bounds are shown in Fig.\ref{fig.bound4}and \ref{fig.bound5}\footnote{These bounds are evaluated using \cite{fernandez2016super} and the MATLAB code therein.}. The proof of Theorem \ref{thm.main} makes essentially use of the bounds on $K_{2D}(\bm{t})$, its partial derivatives and $\bm{Z}$ which is defined in (\ref{eq47}). These bounds are numerically shown in Figs. \ref{fig.bound4}, \ref{fig.bound5} and Fig. \ref{fig.boundZ}, respectively. It may seem surprising at the first sight how the 2-D kernels are drawn versus one variable. Precisely, the proof of Lemma \ref{lemasli2} requires some bounds on 2-D kernels in $\bm{t}:\|\bm{t}\|_2\le 0.212568\lambda_c$. Remark that the positive definiteness of the Hessian matrix $\bm{H}$ in (\ref{hessianmatrix}) is why the radius $0.212568\lambda_c$ is chosen. Since the bounds are monotonic both on $t_1$ and $t_2$ in the interval $0\le t\le 0.212568\lambda_c$ (See Appendix \ref{proofoflemma2} and Fig. \ref{fig.bound1}), it is sufficient to evaluate them on the line $t_1=t_2\le 0.212568\lambda_c$. Moreover, the small grid size $\epsilon=10^{-6}$ that is used in the simulations imposes much computational complexity in the 2-D case. This idea that was first used by \cite{candes2014towards} leads to simple computations. Fig. \ref{fig.bound4} and \ref{fig.bound5} demonstrate bounds on $K_{2D}(\bm{t})$ and its partial derivatives with the condition that $t_1=t_2$.\par 
We further uniformly generate $r=25$ 2-D points in $[0,1]^2$ with coefficients $d_i\sim(0.5+\chi^2(1)) \exp{(j2\pi~{\mathcal{U}[0,1]})}~:i=1,..., r$ and build 2-D Fourier measurements in the form of (\ref{eq2}) up until the square $[-f_c,f_c]^2$ in which $f_c=15$. To reconstruct the location of point sources from the measurements, we first implement the SDP problem (\ref{sdp}) using CVX \cite{grant2008cvx}. Then, the following dual polynomial is obtained by the solution $\hat{\bm{C}}$. 
\begin{align}\label{eq.dualpolynomial}
(\mathcal{F}_{2D}^{*}\hat{\bm{C}})(\bm{t}):=\sum_{\bm{k}\in J}^{}\hat{c}_{\bm{k}} e^{j2\pi \bm{t}^T \bm{k}}.
\end{align}
Based on (\ref{eq.howtofindsupport}) one can localize $|(\mathcal{F}_{2D}^{*}\hat{\bm{C}})(\bm{t})|=1$ to find the signal support as shown in Fig. \ref{fig.dualpoly}.\par
In the last experiment, we evaluate how the success rate scales with changing number of spikes and minimum separation in different number of samples. Fig. \ref{fig.phase} shows that the phase transition occurs near about $1.4/\lambda_c$. 
\section{Conclusion And Future Directions}\label{conclude}
This paper is concerned with the recovery of 2-D point sources where we are given low-resolution measurements caped to integer square $[-f_c,f_c]^2$. We show that TV norm minimization achieves exact recovery when the sources are separated by at least $1.68/f_c$. The proof is based on construction of a 2-D low-pass dual polynomials that can interpolate any sign patterns of the support signal.\par
There are several interesting future directions to be explored. \cite{chi2015compressive} has shown that off-the-grid 2-D point sources can be recovered from partially observed Fourier coefficients as long as the separation is $2.38/f_c$. This bound can be improved to $1.68/f_c$ using the proposed 2-D dual polynomial in Proposition \ref{prop.main}. It is beneficial to consider our 2-D problem in line spectral estimation when the measurements are corrupted with sparse noise using the approach of \cite{fernandez2016demixing}. 
\appendices
\section{Useful Lemmas }\label{usefullemmas}
  The proof sketch of our 2-D low-pass polynomial construction is based on the bounds on the 1-D kernel $K_{\bm{\gamma}}(t)$ in (\ref{eq10}) and its derivatives. These bounds are derived in \cite[Section 4.1]{fernandez2016super} based on Taylor series expansion of Dirichlet kernel and its derivatives around origin.
\begin{lem} \label{lem1}\cite[Lemma 4.4]{fernandez2016super}
For any $\ell \in \{0,1,2,3\}$, if $\tau$ is such that $|f_ct-\tau|\le \epsilon,\quad \forall f_c \geq 10^3 ~\text{and}\quad\forall t\in [0,1]$, then we have the following non-asymptotic bounds on $K_{\bm{\gamma}}^{\ell}(t)$
\begin{align}
\label{eq20}
B_{\gamma ,\ell}^{L}(\tau)-(2\pi)^{\ell+1}f_c^{\ell}\epsilon \le K_{\bm{\gamma}}^{\ell}(t) \le B_{\bm{\gamma} , \ell}^{U}(\tau)+(2\pi)^{\ell+1}f_c^{\ell}\epsilon,
\end{align}
where $B_{\bm{\gamma} ,\ell}^{L}$ and $B_{\bm{\gamma} ,\ell}^{U}$ are defined in \cite[Section B.1]{fernandez2016super}.
\end{lem}
Consequently, one has upper bounds on the magnitude of $K_{\bm{\gamma}}(t)$ and its derivatives as:
    \begin{align}
   \label{eq21}
 |K_{\bm{\gamma}}^{\ell}(t)| \le B_{\bm{\gamma},\ell}^{\infty}(\tau,\epsilon)=\max \{|B_{\bm{\gamma} , \ell}^{L}(\tau)|,|B_{\bm{\gamma} , \ell}^{U}(\tau)|\}\nonumber\\+(2\pi)^{\ell+1}f_c^{\ell}\epsilon.
   \end{align}
\begin{lem}\cite[Lemma 4.6]{fernandez2016super}
	For $f_c \geq 10^3$, $\ell \in \{0,1,2,3\}$ and $ |t|\le \frac{450}{f_c}$ the following decreasing bound is established:
	\begin{align}
	\label{eq22}
	|K_{\bm{\gamma}}^{\ell}(t)| \le b_{\bm{\gamma},\ell}(f_ct),
	\end{align}
\end{lem}
where $b_{\bm{\gamma},\ell}(f_ct)$ is defined in \cite[Section B.2]{fernandez2016super}. Also, there is a global upper bound as:
\begin{align}\label{globalbound}
|K_{\bm{\gamma}}^{\ell}(t)|\le (2\pi f_c)^\ell.
\end{align}
Since the upper bound on sum of $|K_{\bm{\gamma}}^{\ell}(t)|$ and its shift copies are required for 2-D situation, the result of \cite[Lemma 4.7]{fernandez2016super} with minor changes is given below. 
\begin{lem}\label{sumbound}Following \cite[Lemma 4.7]{fernandez2016super},
suppose $0 \in T$ and  $\Delta(T)\ge \tau_{\min}/f_c$ where $\tau_{\min}:=\Delta_{\min}/\lambda_c=1.68$. If $f_c \geq 10^3$ and $\bm{\gamma}=[0.247, 0.339, 0.414]^T$, then for all $t \in [0,\Delta_{min}/2]$ and $\tau$ such that $\tau-\epsilon \le f_ct \le \tau$, $\epsilon \geq 0$, 
\begin{align}
\label{eq23}
\sum_{t_i \in T\setminus\{0\}}|K_{\bm{\gamma}}^{\ell}(t-t_i)| \le H_{\ell}(\tau)+H_{\ell}(-\tau),
\end{align}
where 
\begin{align}
\label{eq24}
H_{\ell}(\tau):=\sum_{i=1}^{20}\max\big\{\underset{u\in \mathcal{G}_{i,\tau}}{\max}(B_{\gamma,\ell}^{\infty}(u,\epsilon), b_{\gamma,\ell}((i+4) \tau_{\min})\big\}+\tilde{C}_{\ell},
\end{align}
where $\mathcal{G}_{i,\tau}$ covers the interval $[i\tau_{\min}-\tau,(i+4)\tau_{\min}]$ with $\epsilon$ equispaced steps and 
\begin{align}
\label{eq25}
\tilde{C}_\ell=\sum_{i=21}^{267}b_{\gamma,\ell}((i-1/2)\tau_{\min})+C_{\ell},
\end{align}
where $C_0=7.89\times
10^{-7}$, $C_1=4.96\times10^{-6}$, $C_2=3.12\times10^{-5}$, and $C_3=1.96\times10^{-4}$. \\The following bounds are beneficial in the proof of Lemmas \ref{lemasli2} and \ref{lemasli3},
\begin{align}
 \label{eq545}
&\sum_{t_i \in T \cap (-\frac{1}{2},0)}\!\!\!\!\!\!\!|K_{\bm{\gamma}}^{\ell}(t-t_i)|\le H_{\ell}(0),\nonumber\\
&\sum_{t_i \in T\cap (0,\frac{1}{2})}\!\!\!\!\!\!\!|K_{\bm{\gamma}}^{\ell}(t-t_i)|\le H_{\ell}(\tau_{\min}/2).
\end{align}
Also, $H_{\ell}(\tau)$ and $H_{\ell}(-\tau)$ are strictly increasing and decreasing functions, respectively.
\end{lem}
\section{Proof of Proposition \ref{prop.main}}  \label{proofprop}
To meet (\ref{signcondition}) and (\ref{diffcondition}) and obtain upper bounds on interpolation coefficients, we present the following Lemma.
\begin{lem}\label{lemmaasli1}
If the condition of Theorem \ref{thm.main} hold, in (\ref{eq13}), there exist coefficient vectors $\bm{\alpha}$, $\bm{\beta}_1$ and $\bm{\beta}_2$ satisfying  
\begin{align}
\label{eq15}
&\|\bm{\alpha}\|_{\infty} \le 1+3.7\times10^{-2},\nonumber\\
&\|\bm{\beta}\|_{\infty} \le 2.4\times10^{-2}\lambda_c,
\end{align}
where $\bm{\beta}=[\bm{\beta}_1^T,\bm{\beta}_2^T]^T$. Also, if $v_1=1$, then
\begin{align}
\label{eq16}
\alpha_{1} \geq 1-3.7\times10^{-2}.
\end{align}
\end{lem}
To control the magnitude of $Q(\bm{t})$ near the elements of the support $T$, we present the following lemmas. Without loss of generality, we can assume that the first element of the support $T$ is located in $\bm{0}$.
\begin{lem}\label{lemasli2}
Assume, without loss of generality, $\bm{0} \in T$. Then if the conditions of Theorem \ref{thm.main} hold, for any $\bm{t}:\bm{0 } < \|\bm{t}\|_2 \le 0.212568\lambda_c$,  $|Q(\bm{t})| < 1$.
\end{lem}
\begin{lem}\label{lemasli3}
Assume, without loss of generality, $\bm{0} \in T$. Then if the conditions of Theorem \ref{thm.main} hold,  $|Q(\bm{t})| < 1$ for any $\bm{t}:  0.212568\lambda_c \le \|\bm{t}\|_2 < \Delta_{\min}$.	
\end{lem}
The proof of our main result requires a numerical upper bound on (\ref{eq20}) and (\ref{twod}) which are shown in Fig. \ref{fig.boundk1k2}.\\ 
\section{Proof of Lemma \ref{lemmaasli1}}
To prove this lemma, the approach of \cite{candes2014towards} is followed. Without loss of generality, we assume the unit square $[0,1]^2$ be mapped to $[-1/2,1/2]^2$. First, (\ref{signcondition}) and (\ref{diffcondition}) are written in matrix form as:
\begin{align}
\label{mat}
\begin{bmatrix}
\bm{E}_{00} & \bm{E}_{10} & \bm{E}_{01}\\
\bm{E}_{10} & \bm{E}_{20} & \bm{E}_{11}\\
\bm{E}_{01} & \bm{E}_{11} & \bm{E}_{02}\\
\end{bmatrix}
\begin{bmatrix}
\bm{\alpha}\\
\bm{\beta}_1\\ 
\bm{\beta}_2
\end{bmatrix}
=
\begin{bmatrix}
\bm{v}\\
\bm{0}\\ 
\bm{0}
\end{bmatrix},
\end{align}
where
  \begin{align}
\label{eq26}
(\bm{E}_{i_1i_2})_{\ell,j}=K_{2D}^{(i_1i_2)}(\bm{t}_\ell-\bm{t}_j).
\end{align}
The interpolation coefficients are calculated from the above equations.ُ Since $K_{\bm{\gamma}}$ and $K_{\bm{\gamma}}^{2}$ are even and $K_{\bm{\gamma}}^1$ is odd, $\bm{E}_{00}$, $\bm{E}_{20}$, $\bm{E}_{11}$ and $\bm{E}_{02}$ are symmetric, while $\bm{E}_{01}$ and $\bm{E}_{10}$ are antisymmetric. Let $\bm{\beta}=[\bm{\beta}_1^T,\bm{\beta}_2^T]^T$, $\tilde{\bm{E}}=[\bm{E}_{10}^T,\bm{E}_{01}^T]^T$ and $\tilde{\bm{E}}_{2}=\begin{bmatrix}
\bm{E}_{20},\bm{E}_{11}\\
\bm{E}_{11},\bm{E}_{02}
\end{bmatrix}$. Therefore, the above matrix system is converted to:
\begin{align}\label{eq.matrixsystem}
\begin{bmatrix}
\bm{E}_{00} & -\tilde{\bm{E}}_{1}^{T}\\
\tilde{\bm{E}}_{1} & \tilde{\bm{E}}_{2} \\
\end{bmatrix}
\begin{bmatrix}
\bm{\alpha}\\
\bm{\beta}\\ 
\end{bmatrix}
=
\begin{bmatrix}
\bm{v}\\
\bm{0}\\ 
\end{bmatrix}.
\end{align}
To bound the infinity norm of the sub-matrices in (\ref{mat}), 1-D results can be used as below:
  \begin{align}
\label{eq27}
&\|\bm{I}-\bm{E}_{00}\|_\infty=\max_{\bm{t}_0}\sum_{\bm{t}_i \in T \setminus \bm{t}_0}|K_{2D}(\bm{t}_i-\bm{t}_{0})|\nonumber\\
&\le\max_{\bm{t}_{0}}\sum_{\bm{t}_i \in T \setminus \bm{t}_{0}}|K_{\bm{\gamma}}({t}_{1i}-t_{10})||K_{\bm{\gamma}}({t}_{2i}-{t}_{20})|,
\end{align}
where $\bm{t}_i=(t_{1i},t_{2i})$, $\bm{t}_0=(t_{10},t_{20})$ and the inequality follows form (\ref{eq12}). To apply the 1-D result as discussed in Appendix\ref{usefullemmas} to 2-D case, we divide the set $T\setminus\{\bm{t}_{0}\}$ into regions $|t_{1i}-t_{01}|\le\Delta_{\min}/2$ or $|t_{2i}-t_{20}|$ $\le \Delta_{\min}/2$ and
$\min(|t_{1j}-t_{10}|,|t_{2j}-t_{20}|) \geq \Delta_{\min}/2$. With this assumption, we reach:
  \begin{align}
\label{eq28}
&\hspace{1cm}\max_{\bm{t}_0}\hspace{-2cm}\sum_{\substack{|t_{1i}-t_{10}|\le \Delta_{\min}/2
~\text{or}~|t_{2i}-t_{20}|\le \Delta_{\min}/2,\\
\bm{t}_{i} \neq \bm{t}_{0}}
}\hspace{-2cm}|K_{\bm{\gamma}}(t_{1i}-t_{10})||K_{\bm{\gamma}}(t_{2i}-t_{20})| \nonumber\\
& \le\max_{t_{20}}~\| K_{\bm{\gamma}}\|_{\infty}\hspace{-0.2cm}\sum_{t_{2j} \neq t_{20}}|K_{\gamma}({t}_{2i}-t_{20})|\nonumber\\
&+\max_{t_{10}}~\| K_{\bm{\gamma}}\|_{\infty}\hspace{-0.2cm}\sum_{t_{1j} \neq t_{10}}|K_{\gamma}({t}_{1i}-t_{10})|\nonumber\\
&\le
 2H_{0}(0)+2H_{0}(0),
\end{align}
where the last inequality stems from (\ref{eq23}) when $\tau=0$ and the fact that $ |K_{\bm{\gamma}}(t)|\le1$. Also the first one is the result of minimum separation $\Delta_{\min}$ between the sources and the union bound.
\\For the last region, we have:
   \begin{align}
 \label{eq29}
 &\hspace{1cm}\max_{\bm{t}_0}\hspace*{-1.5cm}\sum_{\substack{
 		\min(|t_{1i}-t_{10}|,|t_{2i}-t_{20}|) \geq \Delta_{\min}/2, \\
 	\bm{t}_i \neq \bm{t}_0}}\hspace{-1.5cm}|K_{\bm{\gamma}}({t}_{1i}-t_{10})||K_{\bm{\gamma}}(t_{2i}-t_{20})|\nonumber\\
 &\le\bigg(\max_{t_{10}}\hspace{-0.8cm}\sum_{\substack{
 		|t_{1i}-t_{10}| \geq \Delta_{\min}/2,\\t_{1i}\neq t_{10}}}\hspace{-0.8cm}|K_{\bm{\gamma}}(t_{1i}-t_{10})|\bigg)
 	\bigg(\max_{t_{20}}\hspace{-0.8cm}\sum_{\substack{
 			|t_{2i}-t_{20}| \geq \Delta_{\min}/2\\
 		{t_{2i}\neq t_{20},}}}\hspace{-0.8cm}|K_{\bm{\gamma}}(t_{2i}-t_{20})|\bigg)
 	~\nonumber\\&\le (2H_0(0))^2,
 \end{align}
 where the last inequality stems from the minimum separation condition and (\ref{eq23}). Consequently, 
   \begin{align}
 \label{eq30}
&\|\bm{I}-\bm{E}_{00}\|_\infty=\max_{\bm{t}_0}\sum_{\bm{t}_i \in T \setminus \bm{t}_0}|K_{\bm{\gamma}}(\bm{t}_i-\bm{t}_{0})|\\\nonumber\\ & \le 4H_0(0)+4H_0^2(0)\le  3.17\times10^{-2}.
 \end{align}
 By applying the same approach, we reach:
   \begin{align}
 \label{eq31}
 &\|\bm{E}_{10}\|_\infty\le 2H_1(0)+2\|K_{\bm{\gamma}}^{1}\|_{\infty}H_0(0)+4H_1(0)H_0(0)\nonumber\\
 &\le 8.7\times10^{-2}f_c,
\end{align}
where  the bound follows from Fig. \ref{fig.bound1} and (\ref{eq23}). Same bound holds for $\bm{E}_{01}$. Similarly, 
 \begin{align}
\label{eq32}
\|\bm{E}_{11}\|_\infty\le 4\|K_{\bm{\gamma}}^{1}\|_{\infty}H_1(0)+4H_1^2(0) \le  0.181~f_c^2.
\end{align}
Eventually,
\begin{align}
\label{eq33}
\||K^{2}_{\bm{\gamma}}(0)|\bm{I}-\bm{E}_{20}\|_\infty\le 2H_2(0)+2\|K_{\bm{\gamma}}^{2}\|_{\infty}H_0(0)\nonumber\\+4H_2(0)H_0(0) \le 0.583~f_c^2,
\end{align}
where the above inequities come is obtained by (\ref{eq23}) and fact that $\|K_{\bm{\gamma}}^{2}\|_{\infty}$ occur at origin (see Fig. \ref{fig.bound1}).
\\To ease notation, consider,
   \begin{align}
\label{eq34}
&\bm{S}_1=\bm{E}_{20}-\bm{E}_{11}\bm{E}_{02}^{-1}\bm{E}_{11},\nonumber\\
&\bm{S}_2=\bm{E}_{10}-\bm{E}_{11}\bm{E}_{02}^{-1}\bm{E}_{01},\nonumber\\
&\bm{S}_3=\bm{E}_{00}+\bm{S}_2^T\bm{S}_1^{-1}\bm{S}_2-\bm{E}_{01}\bm{E}_{02}^{-1}\bm{E}_{01},
\end{align}
where $\bm{S}_1$ is the Schur's complement of $\bm E_{02}$. Also by the definition of inverse of Schur's complement, we have
\[
\tilde{\bm{E}}_2^{-1}=
\begin{bmatrix}
	\bm{S}_1^{-1}& -\bm{S}_1^{-1}\bm{E}_{11}\bm{E}_{02}^{-1}\\
	-\bm{E}_{02}^{-1}\bm{E}_{11}\bm{S}_{1}^{-1}&\bm{E}_{02}^{-1}+\bm{E}_{02}^{-1}\bm{E}_{11}\bm{S}_{1}^{-1}\bm{E}_{11}\bm{E}_{02}^{-1}\\ 
\end{bmatrix}.
\]
Regarding the fact that $\bm{S}_3$ is the Schur's complement of $\tilde{\bm{E}}_2$, the solution of linear system can be written as
\[
\begin{bmatrix}
\bm{\alpha}\\
\bm{\beta}\\ 
\end{bmatrix}
=
\begin{bmatrix}
\bm{I}\\
-\tilde{\bm{E}}_2^{-1}\tilde{\bm{E}}_1\\ 
\end{bmatrix}
(\bm{E}_{00}+\tilde{\bm{E}}_1^T\tilde{\bm{E}}_2^{-1}\tilde{\bm{E}}_1)^{-1}\bm{v}
\]
\[
\Leftrightarrow
\begin{bmatrix}
\bm{\alpha}\\
\bm{\beta}_1\\
\bm{\beta}_2\\
\end{bmatrix}
=
\begin{bmatrix}
\bm{I}\\
-\bm{S}_{1}^{-1}\bm{S}_{2}\\
\bm{E}_{02}^{-1}(\bm{E}_{11}\bm{S}_1^{-1}\bm{S}_2-\bm{E}_{01}) 
\end{bmatrix}
\bm{S}_3^{-1}\bm{v}.
\]
Respected to $\|\bm{M}^{-1} \|_\infty \le \frac{1}{1-\|\bm{I}-\bm{M}\|_\infty}$ and value of $|K_{\bm{\gamma}}^{2}(0)|$ Fig. \ref{fig.bound1}, we reach: 
\begin{align}
\label{eq35}
\|\bm{E}_{02}^{-1} \|_\infty \le \frac{1}{|\bm{K}_{\gamma}^{2}(0)|-\||\bm{K}_{\gamma}^{2}(0)|\bm{I}-\bm{E}_{02}\|_\infty}\le \frac{0.251}{f_c^2}.
\end{align}
By using $(\ref{eq32})$ and $(\ref{eq33})$,
\begin{align}
\label{eq36}
\||\bm{K}_{\gamma}^{2}(0)|\bm{I}-\bm{S}_{1}\|_\infty \le &\||\bm{K}_{\bm{\gamma}}^{2}(0)|\bm{I}-\bm{E}_{02}\|_\infty\nonumber\\
&+\|\bm{E}_{11}\|^{2}_\infty \|\bm{E}^{-1}_{02}\|_{\infty}\le 0.591~f_c^2.
\end{align}
Similarity $(\ref{eq35})$ reads 
\begin{align}
\label{eq37}
\|\bm{S}_{1}^{-1} \|_\infty \le \frac{1}{|\bm{K}_{\gamma}^{2}(0)|-\||\bm{K}_{\gamma}^{2}(0)|\bm{I}-\bm{S}_{1}\|_\infty}\le \frac{0.251}{f_c^2}.
\end{align}
Next, $(\ref{eq31})$,  $(\ref{eq32})$ and $(\ref{eq35})$  allow to bound $S_2$
\begin{align}
\label{eq38}
&\|\bm{S}_{2} \|_\infty \le \|\bm{E}_{10}\|_\infty+\|\bm{E}_{11}\|_\infty\|\bm{E}_{02}^{-1}\|_\infty\|\bm{E}_{01}\|_\infty\nonumber\\
&\le 9.1\times10^{-2}~f_c,
\end{align}
using $(\ref{eq30})$, $(\ref{eq38})$, $(\ref{eq37})$, $(\ref{eq31})$, and $(\ref{eq35})$ we achieve
\begin{align}
\label{eq39}
\|\bm{I}-\bm{S}_{3} \|_\infty \le \|\bm{I}-\bm{E}_{00}& \|_\infty +\|\bm{S}_2\|_\infty^{2}\|\bm{S}_1^{-1}\|_\infty\nonumber\\
&+\|\bm{E}_{01}\|_\infty^{2}\|\bm{E}_{02}^{-1}\|_\infty\ \le3.6\times10^{-2},
\end{align}
with this bound we have
\begin{align}
\label{eq40}
\|\bm{S}_{3}^{-1} \|_\infty \le \frac{1}{1-\|\bm{I}-\bm{S}_{3}\|_\infty}\le1.037.
\end{align} 
One can bound the interpolation vector with the above results as below:
\begin{align}
\label{eq41}
&\|\alpha\|_\infty\le\|\bm{S}_{3}^{-1} \|_\infty \le1+3.7\times10^{-2},\nonumber\\
&\|\beta_1\|_\infty\le \|\bm{S}^{-1}_1 \bm{S}_2 \bm{S}^{-1}_3\|\le \|\bm{S}^{-1}_1\| \| \bm{S}_2\| \|\bm{S}^{-1}_3\|\le2.4\times10^{-2},\nonumber\\
&\alpha_1=v_1-((\bm{I}-\bm{S}_3^{-1})\bm{v})_1\nonumber\\ &\geq 1-\|\bm{S}_3^{-1}\|_\infty \|\bm{I}-\bm{S}_3\|_\infty \geq 1-3.7\times10^{-2},
\end{align} 
where the last inequality holds when $v_1=1$. The upper bound computations for $\|\bm{\beta}_2\|_\infty$ follows from the same strategy.
\section{Proof of Lemma\ref{lemasli2}}\label{proofoflemma2}
In the previous section we showed that $Q(\bm{t})$ satisfies (\ref{signcondition}) and (\ref{diffcondition}). In this section by showing that the Hessian matrix
\begin{align}\label{hessianmatrix}
\bm{H}=
\begin{bmatrix}
Q^{20}(\bm{t})& Q^{11}(\bm{t})\\
Q^{11}(\bm{t})& Q^{02}(\bm{t})
\end{bmatrix}
\end{align}
is negative definite in the domain $\|\bm{t}\|_2\le 0.212568 \lambda_c$, we prove that the magnitude of $Q(\bm{t})$ can not exceed one in this domain. For this purpose, first using $(\ref{eq20})$ and $(\ref{eq21})$, we establish non-asymptotic bounds on $(\ref{eq12})$ and its partial derivative as follows:
\begin{align}
\label{eq42}
&K_{2D}(\bm{t}) \geq \big[B_{\gamma ,0}^{L}(\tau_1)-(2\pi)\epsilon\big]\big[B_{\gamma ,0}^{L}(\tau_2)-(2\pi)\epsilon\big],\nonumber\\
&K_{2D}^{20}(\bm{t}) \le \big[B_{\gamma ,2}^{U}(\tau_1)+(2\pi)^3f_c^2\epsilon\big]\big[B_{\gamma ,0}^{U}(\tau_2)+(2\pi)\epsilon\big],\nonumber\\
&|K_{2D}^{10}(\bm{t})| \le [B_{\gamma ,1}^{\infty}(\tau_1,\epsilon)] ,\nonumber\\
&|K_{2D}^{11}(\bm{t})| \le \big[B_{\gamma ,1}^{\infty}(\tau_1,\epsilon)\big]\big[B_{\gamma ,1}^{\infty}(\tau_2,\epsilon)\big],\nonumber\\
&|K_{2D}^{21}(\bm{t})| \le  |K^{2}_{\bm{\gamma}}(0)|[B_{\gamma ,1}^{\infty}(\tau_2,\epsilon)],\nonumber\\
&|K_{2D}^{30}(\bm{t})| \le [B_{\gamma ,3}^{\infty}(\tau_1,\epsilon)] ,
\end{align} 
where the functions within the brackets are all monotone in the interval $0\le t \le 0.212568\lambda_c$ as shown in Fig. \ref{fig.bound1}. Regarding this and the fact that $\|\bm{t}\|_{\infty}\le \|\bm{t}\|_2$, we can evaluate the non-asymptotic bounds at $t_1=t_2= 0.212568\lambda_c$ to show that for any $\bm{t}:~\|\bm{t}\|_2\le 0.212568\lambda_c$,
\begin{align}
\label{eq43}
&K_{2D}(\bm{t}) \geq 0.8, &&K_{2D}^{20}(\bm{t}) \le-3.1~f_c^2, \nonumber\\
&|K_{2D}^{10}(\bm{t})| \le0.9f_c, &&|K_{2D}^{11}(\bm{t})| \le0.8~f_c^2, \nonumber\\
&|K_{2D}^{21}(\bm{t})| \le4.09~f_c^2 , &&|K_{2D}^{30}(\bm{t})| \le10.04~f_c^3.
\end{align} 
The same bounds hold for $K_{2D}^{01}$, $K_{2D}^{02}$, $K_{2D}^{12}$ and $K_{2D}^{03}$. Remark that the right-hand side of the second bound in (\ref{eq42}) is multiplication of the negative increasing and the positive decreasing functions $K^{2}_{\bm{\gamma}}$ and $K_{\bm{\gamma}}$, respectively, (See Fig. \ref{fig.bound1}).
\\Later, we leverage the same technique as in \cite{candes2014towards} to bound $\sum_{\bm{t}_i \in T \setminus \bm{0}}|K^{i_1i_2}_{2D}(\bm{t}-\bm{t}_i)|,\quad \forall\|\bm{t}\|_2\le \Delta_{\min}/2$. Consider a special case $i_1=i_2=0$. Without loss of generality, we assume $\bm{t} \in \mathbb{R}_{+}^2$ and $\bm{0} \in T$. Also, we consider that $\bm{t}_i \in \{\min(|t_1|,|t_2|) \geq \Delta_{\min}/2 \} $. To bound the sum, we split the domain $\{\min(|t_1|,|t_2|) \geq \Delta_{\min}/2 \} $ to different regions. First, assume that $\bm{t}_i \in \{\min(|t_1|,|t_2|) \geq \Delta_{\min}/2 \} \setminus \mathbb{R}^{2}_{+} $,
\begin{align}
\label{eq44}
&\sum_{\substack{\bm{t}_i \in \{\min(|t_1|,|t_2|) \geq \Delta_{\min}/2\} \setminus \mathbb{R}_{+}^2, \\ (\|\bm{t}\|_2\le \Delta_{\min}/2)\cap \mathbb{R}^{2}_{+} } }\hspace{-1.8cm}|K_{2D}(\bm{t}-\bm{t}_i)|
	\le\nonumber\\
	& \Bigg(\hspace{-0.1cm}\sum_{\substack{0\le t_1\le \Delta_{\min}/2\\
	t_{1i}\in[\Delta_{\min}/2,\frac{1}{2}]}}\hspace{-0.7cm}|K_{\bm{\gamma}}(t_1-t_{1i})|\Bigg)
\Bigg(\hspace{-0.2cm}\sum_{\substack{0\le t_2\le \Delta_{\min}/2\\
		t_{2i}\in [-\frac{1}{2},-\Delta_{\min}/2]}}\hspace{-0.9cm}|K_{\bm{\gamma}}(t_2-t_{2i})|\Bigg)\nonumber\\
	&+\Bigg(\hspace{-0.2cm}\sum_{\substack{0\le t_1\le \Delta_{\min}/2\\
			t_{1i}\in[-\frac{1}{2},-\Delta_{\min}/2]}}\hspace{-0.9cm}|K_{\bm{\gamma}}(t_1-t_{1i})|\Bigg)
	\Bigg(\hspace{-0.2cm}\sum_{\substack{0\le t_2\le \Delta_{\min}/2\\
			t_{2i}\in [-\frac{1}{2},-\Delta_{\min}/2]}}\hspace{-0.9cm}|K_{\bm{\gamma}}(t_2-t_{2i})|\Bigg)\nonumber\\
		&+\Bigg(\hspace{-0.3cm}\sum_{\substack{0\le t_1\le \Delta_{\min}/2\\
				t_{1i}\in[-\frac{1}{2},-\Delta_{\min}/2]}}\hspace{-0.9cm}|K_{\bm{\gamma}}(t_1-t_{1i})|\Bigg)
		\Bigg(\sum_{\substack{0\le t_2\le \Delta_{\min}/2\\
				t_{2i}\in [-\Delta_{\min}/2,\frac{1}{2}]}}\hspace{-0.8cm}|K_{\bm{\gamma}}(t_2-t_{2i})|\Bigg)\nonumber\\
	&\le H_0(0)H_0(\tau_1)+ H_0(0)H_0(\tau_2)+H_0(0)H_{0}(0)\nonumber\\&\le2 H_0(0)H_0(\|\bm{t}\|_2f_c+\epsilon)+H_0(0)H_{0}(0),
\end{align}
where the first inequality follows from the union bound and splitting the region into three quadrants. The second inequality is a result of the first line of (\ref{eq545}) and Lemma \ref{sumbound}. The last one is obtained using the facts  $|f_ct_i-\tau_i|\le\epsilon~:\forall i=1,2$, $\|\bm{t}\|_{\infty}\le \|\bm{t}\|_2$ and that $H_0(\tau)$ is strictly increasing.\\
 Next, assume that $\bm{t}_i \in \{|t_{1}|\le \Delta_{\min}/2~\text{or}~|t_{2}|\le\Delta_{\min}/2\}$. This leads to 
\begin{align}
\label{eq45}
&\sum_{\substack{\bm{t}_i \in \{|t_{1}|\le \Delta_{\min}/2~\text{or}~|t_{2}|\le\Delta_{\min}/2\}, \\
(\|\bm{t}\|_2\le \Delta_{\min}/2)\cap \mathbb{R}^{2}_{+}}}\hspace{-0.8cm}\hspace{-1cm}|K_{2D}(\bm{t}-\bm{t}_i)|\nonumber\\ &\le \|K_{\gamma}(t)\|_{\infty}\hspace{-0.8cm}\sum_{\substack{0\le t_2\le \Delta_{\min}/2, \\
 t_{2i}\in[-\frac{1}{2},\frac{1}{2}]\setminus \{0\}}} \hspace{-0.5cm}|K_{\bm{\gamma}}(t_2-t_{2i})|\nonumber\\
 &+ \|K_{\gamma}(t)\|_{\infty}\hspace{-0.8cm}\sum_{\substack{0\le t_1\le \Delta_{\min}/2, \\
 t_{1i}\in[-\frac{1}{2},\frac{1}{2}]\setminus \{0\}}} \hspace{-0.6cm}|K_{\bm{\gamma}}(t_1-t_{1i})|\nonumber\\
&\le\hspace{-.5cm}\sum_{\substack{0\le t_1\le \Delta_{\min}/2,\\
		t_{1i}\in[-\frac{1}{2},0)}} \hspace{-.5cm}|K_{\bm{\gamma}}(t_1-t_{1i})|+\hspace{-.5cm}\sum_{\substack{0\le t_2\le \Delta_{\min}/2,\\
		t_{2i}\in[-\frac{1}{2},0)}} \hspace{-.5cm}|K_{\bm{\gamma}}(t_2-t_{2i})|\nonumber\\&+\hspace{-.5cm}\sum_{\substack{0\le t_1\le \Delta_{\min}/2,\\
 t_{1i}\in(0,\frac{1}{2}]}} \hspace{-.5cm}|K_{\bm{\gamma}}(t_1-t_{1i})|+\hspace{-.5cm}\sum_{\substack{0\le t_2\le \Delta_{\min}/2,\\
 t_{2i}\in(0,\frac{1}{2}]}} \hspace{-.5cm}|K_{\bm{\gamma}}(t_2-t_{2i})|\nonumber\\
&
\le 2H_0(0)+2H_0(\|\bm{t}\|_2f_c+\epsilon),
\end{align}
where the first inequality follows from the union bound, minimum separation between point sources and (\ref{eq12}). The second inequality is obtained by splitting $[-\frac{1}{2},\frac{1}{2}]\setminus{0}$ to positive and negative intervals and $|K_\gamma|\le 1$. The last one is obtained by the same approach as in the last inequality in the (\ref{eq44}).\\ 
The only work that remains is to bound the summation when $\bm{t}_i \in \{\min\{|t_1|,|t_2|\} \geq \Delta_{\min}/2 \} \cap \mathbb{R}_{+}^2$. For this purpose, we divide this quadrant to two regions $|t_1-t_{1i}|\le \Delta_{\min}$ or $|t_2-t_{2i}|\le \Delta_{\min}$ and $\min(|t_1-t_{1i}|,|t_2-t_{2i}|)\ge \Delta_{\min}$ as below:
\begin{align}
\label{eq46}
&\sum_{(\|\bm{t}\|_{2}\le \Delta_{\min}/2)\cap\mathbb{R}^{2}_{+}}\hspace{-0.8cm}|K_{2D}(\bm{t}-\bm{t}_i)|\nonumber\\
 &\le\|K_{\gamma}(t)\|_{\infty}\hspace{-0.5cm}\sum_{\substack{0\le t_2\le \Delta{\min}/2, \\t_{2i} \in [\Delta_{\min}/2,\frac{1}{2}]
}} \hspace{-0.5cm}|K_{\bm{\gamma}}(t_2-t_{2i})|\nonumber\\
&+\|K_{\gamma}(t)\|_{\infty}\hspace{-0.5cm}\sum_{\substack{0\le t_1\le \Delta{\min}/2,\\t_{1i} \in [\Delta_{\min}/2,\frac{1}{2}]
		}} \hspace{-0.5cm}|K_{\bm{\gamma}}(t_1-t_{1i})|\nonumber\\
&+\Big(\hspace{-0.4cm}\sum_{\substack{
		0\le t_1\le \Delta_{\min}/2,\\|t_1-t_{1i}|\geq \Delta_{\min},\\ t_{1i}\in [\Delta_{\min}/2,\frac{1}{2}]}}\hspace{-.7cm}|K(t_1-t_{1i})|\Big)\Big(\hspace{-0.4cm}\sum_{\substack{
		0\le t_2\le \Delta_{\min}/2,\\|t_2-t_{2i}|\geq \Delta_{\min},\\ t_{2i}\in [\Delta_{\min}/2,\frac{1}{2}]}}\hspace{-.7cm}|K(t_2-t_{2i})|\Big)\nonumber\\&\le 2H_0(\|\bm{t}\|_2f_c+\epsilon)+H_0^2(\|\bm{t}\|_2f_c+\epsilon),
\end{align}
where the first inequality is obtained using the union bound and minimum separation between the sources. The second one is similar to the last inequality in (\ref{eq44}.\par 
The above approach is applied to $K_{2D}^{i_1i_2}(\bm{t})$ for $\|\bm{t}\|_2\le \Delta_{\min}/2$ as follows:
\begin{align}
\label{eq47}
&\sum_{\bm{t}_i\in T \setminus \bm{0}} |K^{i_1i_2}_{2D}(\bm{t}-\bm{t}_i)| \le \bm{Z}^{i_1i_2}(\|\bm{t}\|_2), 
\end{align}
where for $u >0$,
\begin{align}
\label{eq48}
&\bm{Z}^{i_1i_2}(u) \le\nonumber\\
& H_{i_1}(uf_c+\epsilon)H_{i_2}(0)+H_{i_1}(0)H_{i_2}(0)+H_{i_1}(0)H_{i_2}(uf_c+\epsilon)\nonumber\\
&+\|K^{i_2}\|_{\infty}H_{i_1}(0)+\|K^{i_1}\|_{\infty}H_{i_2}(0)\nonumber\\
&+\|K^{i_2}\|_{\infty}H_{i_1}(uf_c+\epsilon)+\|K^{i_1}\|_{\infty}H_{i_2}(uf_c+\epsilon)\nonumber\\
&+\|K^{i_2}\|_{\infty}H_{i_1}(uf_c+\epsilon)+\|K^{i_1}\|_{\infty}H_{i_2}(uf_c+\epsilon)\nonumber\\
&+H_{i_1}(uf_c+\epsilon)H_{i_2}(uf_c+\epsilon),~~\forall i_1,i_2\in \{0,1,2,3\},
\end{align}
where we used the first line of (\ref{eq545}) and the fact that $H_{\ell}(\tau)$ is strictly increasing. Also, $\|K\|_\infty$ and its derivatives can be obtained using $(\ref{eq21})$ and (\ref{globalbound}) (See Fig. \ref{fig.bound1}). \\To show that $\bm{H}$ is negative definite in $\|\bm{t}\|_2\le 0.212568 \lambda_c$, it is sufficient to have $\text{Tr}(\bm{H})<0$ and $\text{det}(\bm{H})>0$,
 \begin{align}
 \label{eq49}
 &\text{Tr}(\bm{H})=Q^{20}(\bm{t})+Q^{02}(\bm{t}),\nonumber\\
 &\text{det}(\bm{H})=|Q^{20}(\bm{t})||Q^{02}(\bm{t})|-|Q^{11}(\bm{t})|^2.
 \end{align} 
We can easily write $Q^{20}(\bm{t})$ form (\ref{eq13}) as follows: 
\begin{align}
\label{eq50}
Q^{20}(\bm{t})=\sum_{\bm{t}_i \in T}\alpha_iK^{20}_{2D}(\bm{t}-\bm{t}_i)+\nonumber\beta_{1i}K_{2D}^{30}(\bm{t}-\bm{t}_i)\\
+\beta_{2i}K_{2D}^{21}(\bm{t}-\bm{t}_i).
\end{align}
Regarding $(\ref{eq42})$ and $(\ref{eq48})$ we obtain
\begin{align}
\label{eq51}
&Q^{20}(\bm{t}) \le \alpha_1 K^{20}_{2D}(\bm{t})+\|\alpha\|_\infty \hspace{-0.2cm}\sum_{\bm{t}_i\in T \setminus \bm{0}}|K_{2D}^{20}(\bm{t}-\bm{t}_i)|\nonumber\\&+\|\beta\|_\infty \bigg[|K_{2D}^{30}(\bm{t})|+\sum_{\bm{t}_i\in T \setminus \bm{0}}|K_{2D}^{30}(\bm{t}-\bm{t}_i)|+|K_{2D}^{21}(\bm{t})|\nonumber\\&+\sum_{\bm{t}_i\in T \setminus \bm{0}}|K_{2D}^{21}(\bm{t}-\bm{t}_i)|\bigg]\le\alpha_{1}K_{2D}^{20}(\bm{t})+Z^{20}(\|\bm{t}\|_2)\nonumber\\
&+\|\beta\|_\infty \bigg[|K_{2D}^{30}(\bm{t})|+Z^{30}(\|\bm{t}\|_2)+|K_{2D}^{21}(\bm{t})|+Z^{21}(\|\bm{t}\|_2)\bigg]\nonumber\\&\le-1.4809~f_c^2,
\end{align}
where the last inequality is obtained by $Z^{30}(u)$, $Z^{21}(u)$ at $u=0.212568\lambda_c$, $(\ref{eq43})$ which is reported in  Figs. \ref{fig.bound4} and \ref{fig.bound5} evaluated with $\epsilon=10^{-6}$ and the result of Lemma \ref{lemmaasli1}. Further, using $(\ref{eq43})$ and $(\ref{eq47})$, we obtain a bound on $|Q^{11}(\bm{t})|$ at $\|\bm{t}\|_2=0.212568\lambda_c$ as follows:
\begin{align}
\label{eq52}
&|Q^{11}(\bm{t})| \le \|\alpha\|_\infty \bigg[|K_{2D}^{11}(\bm{t})|+Z^{00}(\|\bm{t}\|_2)\bigg]\nonumber\\
&+\|\beta\|_\infty \bigg[|K_{2D}^{21}(\bm{t})|+Z^{21}(\|\bm{t}\|_2)+|K_{2D}^{12}(\bm{t})|+Z^{12}(\|\bm{t}\|_2)\bigg]\nonumber\\
&\le1.4743~f_c^{2}. 
\end{align}
Using $(\ref{eq49})$, $(\ref{eq51}$), and $(\ref{eq52})$, it is clear that $\text{Tr}(\bm{H})<0$ and $\text{det}(\bm{H})>0$ at for any $\bm{t}:~\|\bm{t}\|_2\le0.212568\lambda_c$.\\In the above, we demonstrated that $Q(\bm{t})$ is concave in any small ball around $\bm{0}$. It remains to show that $Q(\mathbf{t}) \le -1$ does not occur in the ball $\|\bm{t}\|_2\le0.212568\lambda_c$ as follow:
\begin{align}
\label{eq53}
Q(\bm{t}) &\geq \alpha_{1} K_{2D}(\bm{t})-\|\alpha\|_\infty Z^{00}(\|\bm{t}\|_2)\nonumber\\
&-\|\beta\|_{\infty}\bigg[|K_{2D}^{01}(\bm{t})|+|K_{2D}^{10}(\bm{t})|+2Z^{01}(\|\bm{t}\|_2)\bigg]\nonumber\\&\geq 0.393.
\end{align}
This concludes the proof.
\section{Proof of lemma \ref{lemasli3}}
\begin{figure}[t]
	\hspace{-.41cm}
	\includegraphics[height=1.95in]{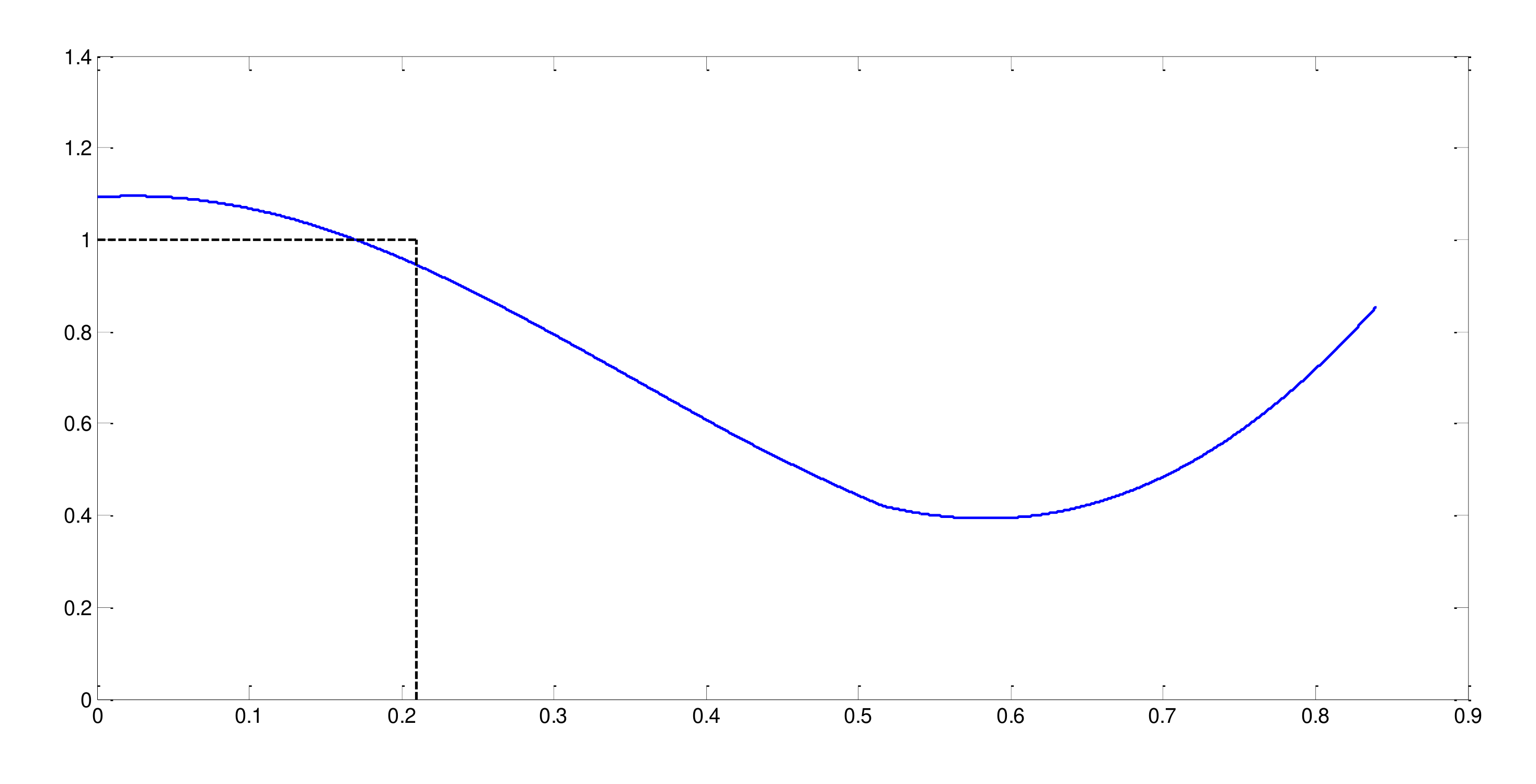}
	\caption{The bound in (\ref{eq590}) versus $\|\bm{t}\|_2/\lambda_c$. The bound is strictly less than one in the region $0.212568\lambda_c\le\|\bm{t}\|_2\le \Delta_{\min}/2$.}\label{fig.boundnew}
\end{figure}

In the previous section using the assumption that $\bm{0} \in T$,  we showed that $|Q(\bm{t})| < 1$ when $\|\bm{t}\|_2\le 0.212568\lambda_c$. In this section, we show that $|Q(\bm{t})|<1$ in $0.212568\lambda_c\le\|\bm{t}\|_2\le \Delta_{\min}$. We first bound $|Q(\bm{t})|$ in $0.212568\lambda_c\le\|\bm{t}\|_2\le \Delta_{\min}/2$ as:
 \begin{align}
\label{eq590}
&|Q(\bm{t})|=\bigg|\sum_{\bm{t}_i \in T}\alpha_iK_{2D}(\bm{t}-\bm{t}_i)+
\beta_{1i}K_{2D}^{10}(\bm{t}-\bm{t}_i)
\nonumber\\
&+\beta_{2i}K_{2D}^{01}(\bm{t}-\bm{t}_i)\bigg|\le
 \|\bm{\alpha}\|_{\infty}\bigg[|K_{2D}(\bm{t})| \nonumber\\
 &+\hspace{-0.3cm}\sum_{\bm{t}_i\in T\setminus{\bm{0}}}\hspace{-0.3cm}|K_{2D}(\bm{t}-\bm{t}_i)|\bigg]+\hspace{-.1cm}\|\bm{\beta}\|_{\infty}\bigg[|K^{10}_{2D}(\bm{t})|\hspace{-.1cm}+\hspace{-.1cm}|K^{01}_{2D}(\bm{t})|\nonumber\\&+\hspace{-0.3cm}\sum_{\bm{t}_i\in T\setminus{\bm{0}}}\hspace{-0.3cm}|K^{10}_{2D}(\bm{t}-\bm{t}_i)|+\hspace{-0.3cm}\sum_{\bm{t}_i\in T\setminus{\bm{0}}}\hspace{-0.3cm}|K^{01}_{2D}(\bm{t}-\bm{t}_i)|\bigg]\nonumber\\
 &\le\|\bm{\alpha}\|_{\infty}\bigg[|K_{\bm{\gamma}}(\|\bm{t}\|_{2})|+Z^{00}(\|\bm{t}\|_{2})\bigg]
+\|\bm{\beta}\|_\infty\bigg[|K^{1}_{\bm{\gamma}}(\|\bm{t}\|_{2})|\nonumber\\&+|K^{1}_{\bm{\gamma}}(\|\bm{t}\|_{2})|+2Z^{10}(\|\bm{t}\|_2)\bigg],\nonumber\\ 
\end{align}
where in the last inequity the property $|K_{2D}(\bm{t})|\le|K_{\bm{\gamma}}(t_1)||K_{\bm{\gamma}}(t_2)|\le |K_{\bm{\gamma}}(\|\bm{t}\|_2)|$ is used, since $K_{\bm{\gamma}}(t_1)$ and $K_{\bm{\gamma}}(t_2)$ are both decreasing in the aforementioned region \footnote{See Fig. \ref{fig.bound1}.} and the maximum is achieved at $t_1=0, t_2=\|\bm{t}\|_2$, or vice versa. $|K^{01}_{2D}(\bm{t})|$ and $|K^{10}_{2D}(\bm{t})|$ can be bounded with similar approach. We numerically show that the upper bound on $|Q(\bm{t})|$ is less than one in the region $0.212568\lambda_c\le\|\bm{t}\|_2\le \Delta_{\min}/2$ as in Fig. \ref{fig.boundnew}.
\\
Assume that $\bm{t}_c$ is the closest point in the support to $\bm{0}$. It might be possible that $\Delta_{\min}\le\|\bm{t}_c\|_2<2\Delta_{\min}$. Therefore, $|Q(\bm{t})|$ must be bounded in the region $\Delta_{\min}/2\le\|\bm{t}\|_2< \Delta_{\min}$. For this purpose, we extend the  approach in \cite[Lemma 4.3]{fernandez2016super} to two-dimensional situation as below:
 \begin{align}
\label{eq56}
|Q(\bm{t})| \le& \|\alpha\|_\infty\bigg[|K_{2D}(\bm{t})|+Z^{00}_{\infty}\bigg]\nonumber\\
&+\|\beta\|_\infty\bigg[|K_{2D}^{01}(\bm{t})|+|K_{2D}^{10}(\bm{t})|+2Z^{10}_{\infty}\bigg].
\end{align} 
where the upper bounds on $|K_{2D}(\bm{t})|$, $|K^{01}_{2D}(\bm{t})|$ and $|K^{10}_{2D}(\bm{t})|$ in $\Delta_{\min}/2\le\|\bm{t}\|_2< \Delta_{\min}$  can be obtained by $(\ref{eq21})$ and $(\ref{eq22})$ as:
  \begin{align}
\label{eq54}
&|K_{2D}(\bm{t})|\le |K_{\bm{\gamma}}(t_1)||K_{\bm{\gamma}}(t_2)|\nonumber\\
& \le \max_{\substack{\Delta_{\min}/2\le t_1< \Delta_{\min}}} \!\!\!\!\!\!\!|K_{\bm{\gamma}}(t_1)|\nonumber\\
&\le\bigg[\max \bigg\{\underset{u \in \mathcal{G}}{\max} ~\big(B_{\bm{\gamma},0}^{\infty}(u,\epsilon), b_{\bm{\gamma},0}(\tau_{\min})\big) \bigg\}\bigg]=0.152,\nonumber\\
&|K_{2D}^{01}(\bm{t})|\le |K_{\bm{\gamma}}(t_1)||K^{1}_{\bm{\gamma}}(t_2)|\nonumber\\
&\le \max_{\substack{\Delta_{\min}/2\le t_2< \Delta_{\min}}}\!\!\!\!\!\!\!|K^{1}_{\bm{\gamma}}(t_2)|\nonumber\\
& \le\bigg[\max \bigg\{\underset{u \in \mathcal{G}}{\max} ~\big(B_{\bm{\gamma},1}^{\infty}(u,\epsilon), b_{\bm{\gamma},1}(\tau_{\min})\big) \bigg\}\bigg]=0.825,
\end{align}
where $|K_{\bm{\gamma}}| \le 1$ is used and the upper bounds are obtained by searching for maximum value of the kernel in the interval $[\Delta_{\min}/2, \Delta_{\min}]$ with a grid step size $\epsilon=10^{-6}$. Similar to $|K^{01}_{2D}|$ the same bound holds for $|K^{10}_{2D}|$ as well. Also, $Z_{00}^{\infty}=0.66$ and $Z_{10}^{\infty}=2.2919$ are obtained using $(\ref{eq48})$ and the second line of (\ref{eq545}).
To complete the proof, we have numerically shown $|Q(\bm{t})|\le 0.9866$ in the domain $\Delta_{\min}/2 \le \|\bm{t}\|_2 < \Delta_{\min}$ with step-size $\epsilon=10^{-6}$. This concludes the proof.  
\bibliographystyle{ieeetr}
 \bibliography{mypaperbibe}
\end{document}